# Investigating the formation and growth of Titan's atmospheric aerosols using an experimental approach

Zoé Perrin[1], Nathalie Carrasco[1,2], Thomas Gautier[1,3], Nathalie Ruscassier[4], Julien Maillard[5], Carlos Afonso[5] and Ludovic Vettier[1]

1 : LATMOS, Université Paris-Saclay, UVSQ, CNRS, 78280 Guyancourt, France
2 : Ecole Normale Supérieure Paris-Saclay, 91190 Gif sur Yvette, France
3 : LESIA, Observatoire de Paris, Université PSL, CNRS, Sorbonne Université, 92195 Meudon, France
4 : LGPM, CentraleSupélec, Université Paris-Saclay, 91190 Gif-sur-Yvette, France
5 : COBRA, Université de Rouen, UMR 6014 & FR 3038, IRCOF, 76821 Mont St Aignan Cedex, France

**Abstract:**

Titan has a climate system with similarities to Earth, including the presence of a thick atmosphere made up of several atmospheric layers. As on Earth, Titan's climate is influenced by several factors: the gaseous species making up the atmosphere, the energy deposited on the satellite, and solid organic aerosols. Indeed, numerous observations have revealed the presence of solid particles in the form of an opaque orange haze in Titan's atmosphere, influencing radiation balance and atmospheric dynamics, for example. However, the influence of these suspended solid particles seems to evolve according to the atmospheric altitude where they are located, certainly testifying to the presence of organic solids with different physico-chemical properties. At present, it is suspected that several populations/classes of atmospheric aerosols may form following different chemical pathways, and that aerosols undergo growth processes that modify their properties. In this experimental study, we propose new observations on the evolution of morphological and chemical properties observed on Titan aerosol analogues, produced from a mixture of 20% $CH_4$ and 80% $N_2$ injected into a dusty RF plasma experiment. Using SEM (morphological) and FTICR-LDI-MS (chemical composition) analyses, we observe that properties evolve according to certain formation and growth mechanisms, which differentiate over time. The evolution of the neutral gaseous chemical composition analyzed in-situ by QMS in parallel shows correlations with the evolution of solid properties, testifying to the selective involvement of certain neutral products in the formation and growth mechanisms of solid aerosols. By linking the analyses of the gas phase and organic solids, we propose the calculation of an uptake coefficient between six neutral gaseous products ($C_2H_2$, HCN, $C_2H_6$, $C_2H_3N$, $HC_3N$, $C_2N_2$) and the surface of the Titan aerosol analogues produced in this study.

**Keywords:** Titan, aerosol analogues, laboratory experiments, uptake coefficient

# I - Introduction:

Saturn's largest satellite, Titan, has many features similar to those of the Earth, including the presence of a thick atmosphere containing mostly the same gaseous compound, dinitrogen $N_2$ (Lindal et al., 1983; Broadfoot et al., 1981). However, the second major gas on Titan is methane $CH_4$ (Samuelson et al., 1997; Hanel, 1982). With the help of solar photons or energetic particles from Saturn deposited in its atmosphere, the combined photolysis of $N_2$ and $CH_4$ sets up endogenous organic chemistry. Observations made by the Voyager I and II, and Cassini-Huygens space missions have provided a wealth of information on the atmospheric processes taking place on the satellite.

Atmospheric organic chemistry has been shown not only to produce a wide range of gaseous products, but also to form solid aerosols in the form of an opaque orange haze. These organic aerosols appear as high as the upper atmosphere (Waite et al., 2007), then are suspected to sediment slowly towards the surface (Lavvas et al., 2011). Indeed, observations made during the Cassini-Huygens mission have made it possible to calculate the density profiles of aerosols present from Titan's upper to lower atmosphere (Lavvas et al., 2009; Bellucci et al., 2009; Vinatier et al., 2010.b). Different altitude-restricted regions stand out with higher aerosol quantities, including a main layer in the stratosphere where aerosol density is considered highest. Cassini's Visible and Infrared Mapping Spectrometer (VIMS) has observed that atmospheric aerosols ultimately cover part of Titan's surface (Tomasko et al., 2005; Clark et al., 2010; McCord et al., 2006).

The physical and chemical properties of Titan's atmospheric aerosols have been progressively elucidated thanks to numerous observations made during space missions, laboratory experiments reproducing the formation of Titan-like aerosols, and numerical models. These properties confirm that Titan's aerosols have a strong influence on the atmospheric processes regulating climate. However, the involvement of aerosols is complex to understand, as a variation in their effects has been observed, possibly reflecting the presence of aerosols with different properties, morphological or chemical for example. Between the atmospheric region where the particles are formed and Titan's surface, environmental conditions will vary, and certainly produce different populations/classes of organic aerosols (Trainer et al., 2004; Lorentz et al., 2001), or even induce growth of solids leading to modification of their own physico-chemical properties (Lavvas et al., 2011).

Aerosol production on Titan takes place predominantly in the ionosphere/thermosphere at altitudes between 1,000 and 650 km (Waite et al., 2007; 2009; Lavvas et al., 2011). This atmospheric layer is characterized by a low-pressure (below $10^{-4}$ - $10^{-5}$ mbar) and low-temperature (below 140 K) environment. In this environment, depending on altitude, latitude $^{and}/_{or}$ longitude, various energy sources are deposited, such as solar radiation in the ultraviolet (photons and photoelectrons) and magnetospheric electrons from Saturn (Sittler et al., 2009; 2010; Lavvas et al., 2011).

Subsequently, still under the effect of deposited energy, the photolysis of many of the gaseous products detected is likely to form precursors, which can take part in polymerization reactions leading to the formation of increasingly complex polymers. These polymerization reactions are considered to proceed until the formation of polymers reaching masses of between 600 and 2000 amu, marking the transition from polymer nucleation to solids, i.e. the appearance of a solid aerosol (Lavvas et al., 2008.a).

Numerical models (Lebonnois et al., 2002; Wilson and Atreya, 2003; Lavvas et al., 2008.a; 2008.b; Krasnopolski, 2009) have reproduced the possible chemical pathways leading to gas-to-solid transformation in the observed context of Titan's organic chemistry. One of the preferred scenarios for the formation of organic solids is the polymerization of polycyclic aromatic molecules, involving as nucleation nucleus the simplest gaseous aromatic hydrocarbon detected in Titan's ionosphere, benzene $C_6H_6$ (Vuitton et al., 2007; Waite et al., 2007). After several flybys of Titan by the Cassini space mission, measurements by the INMS instrument show a higher abundance of $C_6H_6$ ($10^{-6}$ - $10^{-7}$; ppm) at around 1000 km altitude, compared with those observed in the lowest atmospheric layers (Waite et al., 2009). In this atmospheric region, where $C_6H_6$ production is favored, polymer growth in the form of polycyclic aromatics could also be favoured (Wilson and Atreya, 2003). It has been shown that PAH polycyclic aromatic hydrocarbons can be formed from reactions combining $C_6H_6$ with other carbon species detected in Titan's ionosphere, such as $C_2H_2$ acetylene and $C_6H_5$ phenyl radical (Bauschicher and Ricca, 2000; Lavvas et al., 2008.a). It has also been revealed that photolysis of $C_6H_6$ in nitrogen-based gas mixture can form polycyclic aromatic constituents containing nitrogen. Nitrogen is indeed chemically available in the upper atmosphere, in both atomic (fundamental $N(^4S)$ and excited $N(^2D)$, $N(^2P)$, ect) and ionic ($N^+$, $N_2^+$) forms, but also incorporated into neutral gaseous products such as nitriles like hydrogen cyanide HCN and cyanoacetylene $HC_3N$, radical products such as amines $-NH_x$, as well as ionic products (Waite et al., 2007; Lavvas et al., 2008.a).

Experimentally, it has been shown that photolysis or irradiation of a gas mixture involving particularly that of $C_6H_6$ can form solid particles from reactions producing polymers in the form of polycyclic aromatics (Trainer et al., 2013; Yoon et al., 2014; Gautier et al., 2017; Sciamma O'-Brien et al., 2017). It has been observed that low temperature does not prevent these reactions (Ricca et al., 2001; Sciamma O'-Brien et al., 2017), and that they can be initiated by energy ranges similar to that deposited on Titan (minimal to maximal). The rate of reactions involving $C_6H_6$ exhibits pressure independence, however the size of polymers formed from these reactions depends on pressure and temperature, becoming increasingly larger as pressure increases and temperature decreases (Wang and Frencklach, 1994; Park et al., 1999). The presence of nitrogen in its ionic $N_2^+$ form can promote gaseous chemical processes involving gaseous hydrocarbons, such as $C_6H_6$ gas formation and PAH polymerization (Imanaka and Smith, 2007; 2010).

However, laboratory experiments have observed that polymerization involving $C_6H_6$ as the nucleation core shows low incorporation of nitrogen into the solid, associating preferentially with hydrogen under low-pressure conditions and when nitrogen-containing gaseous moieties and products are scarce in the gas phase (Trainer et al., 2013; Yoon et al., 2014). By increasing the presence of nitrogenous gaseous species (H remaining dominant), it has been shown that nitrogen can be more easily fixed as $-NH_x$ and $-CN$ forms in polycyclic aromatic polymers (Gautier et al., 2017). Under low-pressure conditions, nitrogen can participate dominantly as a copolymer via radicals or neutral species such as HCN, acetonitrile $C_2H_3N$ or $HC_3N$, but where the presence of fragments such as $N(^4S)$ is essential with high production kinetics (high photolysis energy; Imanaka and Smith, 2010).

In parallel with the presence of $C_6H_6$ in Titan's ionosphere, lighter gaseous species such as $C_2H_2$, ethylene $C_2H_4$, ethane $C_2H_6$ and HCN have been detected and considered as abundantly formed primary gaseous products (Waite et al., 2007; 2009; Magee et al., 2009; Kranopolsky, 2009; Carrasco et al., 2008; Lavvas et al. 2008.a; De La Haye et al., 2008; Vuitton et al., 2007; Vuitton et al., 2007; Wilson and Atreya, 2004; Toublanc et al., 1995; Lara et al., 1996; Yung et al., 1984). Their presence is suspected of participating in the synthesis of gaseous aromatic molecules such as $C_6H_6$, but also of being directly involved in reactions leading to the formation of polymers with a rather linear or branched (aliphatic) structure up to the production of organic solids (Lebonnois et al., 2002; Wilson and Atreya, 2003; Lavvas et al. 2008.a).

It has been shown that $C_2H_2$ photolysis products can form copolymers by combining with radicals such as CN, for example (Allen et al., 1980; Opansky and Leone, 1996.a; Chastaing et al., 1998; Vakhtin et al., 2001.a). Other studies have presented that neutral products in the form of nitriles such as HCN and $HC_3N$ can also be considered as reactants in copolymerization processes involving radicals (Thompson and Sagan, 1989). However, it has been shown that reactions involving the CN radical are faster when it associates with gaseous hydrocarbons than with nitriles (Seki and Arabe, 1993; Butterfield et al., 1993). The rate of reaction between the CN radical and gaseous nitriles is also considered to be pressure-dependent (Wang and Frenklach, 1994). Other chemical routes producing pure polymers of the gaseous product considered reactive have also been proposed (Allen et al., 1980; Rettig et al., 1992; Clark and Ferry, 1997). However, it has been shown that the optical properties of pure poly-HCN structures do not exhibit the correct wavelength dependence deduced from Titan's geometric albedo (Khare et al., 1994).

In correlation, experimental observations have also shown that organic solid formation mechanisms involving greater nitrogen participation can also take place with non-aromatic gaseous species such as $C_2H_2$, $C_2H_4$ and $C_2H_6$ (Imanaka et al., 2004; Carrasco et al., 2009; 2012; Sciamma O'-Brien et al., 2010; 2014; 2017; Hörst et al., 2018; Gautier et al., 2011; 2012; 2014; the present study). It has been observed that low-temperature environmental conditions do not prevent these reactions (Sciamma O'-Brien et al., 2014; 2017), and that they can take place with energies equivalent to the range where solar flux is maximum at 1000 km in Titan's atmosphere. These reactions involving light hydrocarbons such as $C_2H_2$ and $C_2H_4$, can generate solids by forming aliphatic compounds, where nitrogen seems to be more easily incorporated into a branched chain structure, via nitrogen-containing copolymers such as radicals (CN, NH, HCCN), neutral species ($HC_3N$, HCN) or ionic species (Sciamma O'-Brien et al., 2010; Gautier et al., 2014). The presence of hydrogen can also influence these processes, saturating the surface of forming solids and maintaining solid constituents in aliphatic form, or it has been observed that it can bind readily to nitrogen under low-pressure conditions (Sekine et al., 2008.a; 2008.b; Sciamma O'-Brien et al., 2010; Carrasco et al., 2012).

It is also suspected that the morphology of solid aerosols is influenced by these processes. Cassini's high-altitude observations predict that the first aerosols formed are spherical in shape, and characterize the lower limit considered for aerosol size. Based on the largest size detected within the anions present in the ionosphere, Coates et al. (2007) estimated a mean diameter for these first spherical monomers equivalent to 6 nm. In their numerical model coupling photochemistry with aerosol production for Titan's atmosphere,

Lavvas et al. (2008.a) set an average diameter of less than a nanometer (0.735 nm) for these same monomers.

Laboratory experiments have shown that the overall morphology of aerosol analogues resulting from polymerization processes of PAHs (with or without nitrogen) or aliphatics (often dominated by nitrogen), appears similar with a quasi-spherical shape (Hadamcik et al., 2009; Hörst and Tolbert, 2013; Trainer et al., 2013; Sciamma O'-Brien et al., 2017). The average diameter observed on different laboratory analogues can vary from 10 to 500 nm, of which the following effects have been observed on numerous occasions: the photolyzed amount of $C_6H_6$ and $CH_4$ gas, i.e. the presence of gaseous hydrocarbons, favors the increase in volume of spherical particles as well as their productions in number.

Following their formation, organic aerosols are expected to travel through the various atmospheric layers present on Titan. The numerical model of Lavvas et al. (2011) predicts that once primary solid aerosols have formed, they can begin growth processes that allow their physico-chemical properties to evolve. Lavvas et al. (2011) consider that during the evolution of aerosols, mechanisms linked to chemical and microphysical processes intervene, and will initiate different morphological growth. Depending on the solid growth environment considered, the involvement of a chemical process may become predominant over a microphysical one, and vice versa.

Complementing what has been observed so far, we present ~~propose~~ in this work an experimental study providing new observations of the evolution of morphological and chemical properties of Titan analog aerosols produced using the PAMPRE dusty plasma reactor (Szopa et al., 2006). We observe a microphysical and chemical temporal evolution of solid aerosols correlating with the observed temporal variation in the kinetics of certain neutral gaseous products formed within the experimental simulation. By combining the different analyses, we were able to dissociate the different stages of formation and growth that Titan's analog aerosols undergo over time. These formation and growth mechanisms appear to involve various neutral gaseous molecules, but also the solid monomers/aerosols themselves.

## II - Materials and Methods:

## II.a - Experimental simulations: Samples production

The RF dusty plasma experiment named PAMPRE (French acronym for "Production d'Aérosols en Microgravité par Plasma REactif") has demonstrated its effectiveness in producing solid analogues to Titan's atmospheric aerosols, simulating the organic chemistry observed on the satellite. By injecting a gaseous $N_2$ - $CH_4$ mixture into this open-flow reactor, it has been shown that once the plasma discharge is triggered, the fragmentation of $CH_4$ combined with that of $N_2$ enables the formation of numerous gaseous products. Their proportions will initially vary through time during a gaseous kinetic regime called transient, before becoming constant over time characterizing a stable kinetic regime (Sciamma O-Brien et al., 2010; Wattieaux et al., 2015; Gautier et al., 2011). In the first instants of the discharge when gas proportions vary, it has been shown that suspended solid aerosols also appear within the experimental plasma (Alcouffe et al., 2010; Wattieaux et al., 2015). Then it is strongly suspected, when the gas composition is stable, that the solids are ejected outside the plasma, where their physical and chemical properties are considered no longer to evolve/change. By varying the initial proportion of $CH_4$ from 1 to 10%, several studies have shown that during the transient gas kinetic phase, the proportion of $CH_4$ consumed and the duration of this consumption also vary, inducing different final chemical composition of the gas phase and of the solid aerosols, and in particular different nitrogen contents (Sciamma O-Brien et al., 2010; Gautier et al., 2011; 2014; Carrasco et al., 2012). It has also been observed that the proportion of $CH_4$ consumed and its consumption duration can be regulated using the gas injection flow, where for example for a fixed initial proportion of 5%, decreasing the gas flow from 55 to 10 sccm increases the duration and quantity of $CH_4$ consumed. As a consequence, aerosols formed at a gas flow of 10 sccm have a micrometric size compared with the 400 - 500 nm diameter observed with aerosols produced at 55 sccm (Sciamma O-Brien et al., 2010; Perrin et al., 2021; Hadamcik et al., 2009).

Depending on their abundance, some gaseous products are suspected to play a role in the formation and growth of the solid aerosols in the PAMPRE experiments. In addition, it has been shown that certain forces applied to each individual solid particle levitating in the plasma, notably the drag force of the gases (controlled by gas injection flow) and the weight of the solid particle, have a major influence in breaking the force balance equilibrium that causes aerosols to be ejected from the plasma. These forces influence the aerosol exposure time and the gas-solid interaction in the PAMPRE plasma. When the RF generator regime of the PAMPRE experiment operates continuously, the exact residence time of the aerosols in the plasma is not known. However it can be controlled when the plasma experiment is operated with alternating current and with control of the shape of the signal using a function generator, enabling regular and instantaneous alternation of ON and OFF periods of the plasma discharge with a precise time duration (Alcouffe et al., 2010). With this generator mode named "pulsed", we constrain the exposure time of solid aerosols. Indeed they will reside in the plasma for the duration set at the ON period, and as soon as the discharge is switched off, the aerosols will sediment out of the plasma (Hadamcik et al., 2009).

In this study, we carried out seven different productions of Titan aerosol analogues, varying the mode of operation of the RF generator of the PAMPRE dusty plasma reactor. Similar to previous studies, throughout each production run: the pressure is stabilized at 0.9 mbar; an RF discharge with an incident power of 30 W and a frequency of 13.56 MHz is triggered between the two electrodes of the plasma cage. As shown by Alcouffe et al. (2008), gas heating is negligible and its temperature can be estimated at around 300 K.

The gas injection flow is set at 2.5 sccm and regulated by a mass flow controller (MKS) and a primary pumping system. For the initial gas mixture, proportions of 80% $N_2$ and 20% $CH_4$ are chosen. Increasing the initial proportion of $CH_4$ from the 1-10% observed in Titan's atmosphere, and reducing the gas flow from the usual 55 sccm of the PAMPRE reactor, enable to consume a significant amount of $CH_4$, and thus promote the chemical processes leading to the formation of the gaseous products (details in section III.a.2). Although an initial proportion of 20% $CH_4$ is no longer representative of Titan's atmospheric conditions, the section III.a.1 details why we still consider that the experimental conditions of this study can represent the satellite's atmospheric organic chemistry.

Six productions of aerosol analogues were carried out using the RF generator in pulsed mode, where for each production a different duration was chosen for the ON period of the discharge (summarized in table 1), allowing the stop the solid growth at different times during the transient regime. The shortest duration chosen here is 20s, as no organic films on the plasma walls or solid aerosols were produced for shorter durations. When producing the aerosol analogues, the duration of an ON period is repeated a large number of times to collect a sufficient quantity of solid for analysis. Aerosol analogues production achieved with the longest controlled discharge (Pul145s) corresponds to the time at which gaseous chemistry begins to stabilize under the experimental conditions set in this study (around 150 seconds;

described section II.a.2). For all productions, between each ON period, the time during which the discharge is switched off is constant and fixed at 10 minutes, corresponding to the time required for complete gas renewal in the PAMPRE reactor at an gas flow of 2.5 sccm (gas transport from the injection inlet to the pumping outlet inside the chamber). A single production of aerosol analogues was carried out, this time with the RF generator running continuously, i.e. without interrupting the plasma discharge (Continuous production; table 1). The experimental parameters defined for the seven aerosol analogues productions are summarized in table 1. Although we could not identify the reason why, the production rate for Pul50s probably corresponds to an outlier result as it is very unlikely this production could have a production rate almost 200 times the one of Pul35s and 8 times the one of Pul75s.

**Table 1 -** Experimental parameters set in the PAMPRE dusty plasma reactor for the various Titan's aerosol analogues productions realized in this study.
*During production, as aerosols are ejected from the plasma, they sediment in a glass vessel surrounding the plasma cage, where they are collected and their mass is measured using a precision balance (correspond of $m_{Tot,harvested}$ of Eq. (5) in section III.b). We note that inside the PAMPRE reactor, very few solid matter was found outside the glass vessel.
** correspond of $t_{Tot,Prod}$ of Eq. (4).

| Experimental conditions | | Titan analog aerosol production | | | | |
|---|---|---|---|---|---|---|
| Injected gas mixture | Gas injection flow | Names | Duration of RF discharge (s) | Total production time** | Mass of aerosol analogues collected* (± 0.1 mg) | Production rate (mg.h$^{-1}$) |
| 80% N$_2$ 20% CH$_4$ | total gas flow: 2.5 sccm with - N$_2$ at 2 sccm - CH$_4$ at 0.5 sccm | Pul20s | 20 | 91h47min | 0.4 | 4.3.10$^{-3}$ |
| | | Pul35s | 35 | 8h | 1.0 | 1.3.10$^{-2}$ |
| | | Pul50s | 50 | 6h | 14.2 | 2.4 |
| | | Pul75s | 75 | 14h25min | 4.4 | 0.3 |
| | | Pul100s | 100 | 8h24min | 6.2 | 0.7 |
| | | Pul145s | 145 | 16h04min | 24.2 | 1.5 |
| | | Continuous | continue | 49h | 120.0 | 2.4 |

## II.b - Gas phase analysis method: In-situ mass spectrometry

We carried out a mass spectrometry analysis of the gaseous chemical composition during the Continuous production of aerosol analogues (table 1). The gaseous molecules within the plasma are pumped through a 150 μm hole into a quadrupole mass spectrometer (QMS EQP-300, Hiden Analytical). In the spectrometer's ionization chamber (ionization source EQP: Integral Electron Impact Ionizer), the gases are ionized by an electron beam with an energy of 70 eV, emitted by an oxide-coated iridium filament with emission current at 0.2 μA. The ions formed in the ionization chamber are then separated according to their mass-to-charge ratio ($m/z$), by a fast neutral energy analyzer (EQP Analyzer: 45° Parallel Plate Electrostatic Energy Analyzer). The analyzer's abundance sensitivity is 0.1 ppm. For each $m/z$ transmitted, an SEM (Secondary Electron Multiplier) detector converts the ions impacting the detector into a measured current that can be converted into c/s (counts/second). When not in acquisition mode, the mass spectrometer is continuously pumped in secondary vacuum at a limiting pressure of $10^{-9}$ mbar, using a turbomolecular pump (HiPACE 300, Pfeiffer-Vacuum).

In this study, the QMS signal is taken in two modes: the RGA (Residual Gas Analysis) mode and the MID (Multiple Ion Detection) mode. In RGA mode, the m/z ratios detected are represented by bargraphs, and measured in the m/z range from 1 to 100. In MID mode, the intensity of a fixed m/z ratio is monitored over time. For each acquisition mode, the time of the detector spent on each measurement point is fixed at 200 ms (dwell time). To ensure that no secondary reactions take place in the mass spectrometer's ionization chamber that could alter the composition of the gas mixture, the pressure within the chamber does not exceed $10^{-6}$ mbar.

## II.c - Method for analyzing the morphology of aerosol analogues: Scanning electron microscopy (SEM)

The morphology of the seven Titan's aerosol analogues productions is observed using a FEG-SEM field emission gun scanning electron microscope (LEO GEMINI 1530 SEM, Zeiss), at the LGPM/LMPS laboratory (Centralesupélec, University of Paris-Saclay, France).

For the measurements, the aerosol analogues considered as an insulating organic material are first deposited on an aluminum grid (1cm of diameter) with a very fine mesh (0.038/0.038 mm), which are then arranged within a metallizer (Quorum Q150T ES) depositing a very thin layer of chromium or tungsten on the surface of the grids containing aerosol analogues. In the SEM chamber, measurements are carried out in a secondary vacuum (less than $10^{-7}$ mbar), after cleaning the samples with nitrogen plasma. The instrument was used with the following technical features: high-resolution imaging mode ranging from 1 to 3 nm; an In-Lens electronic detector. Observation conditions were as follows: acceleration voltage 2 kV; diaphragm size 30 µm; working distance 4 to 6 mm (WD); observation magnification variation from x5K to x80K.

Using the SEM images and a processing code (Matlab), the spherical shape of the particle (see figure 4) enables their detection as perfect circles using a Hough transform, and to determine the diameter distribution for each production of aerosol analogues. To improve the detection of the particles spherical curvature, various contrast treatments were applied to the SEM images using the "Image Processing Toolbox$^{TM}$ " library of Matlab.

## II.d - Method for analyzing the chemical composition of aerosol analogues: high-resolution mass spectrometry

The chemical composition of the seven aerosol analogues productions was analyzed by FTICR-LDI-MS Fourier transform ion cyclotron resonance (Solarix XR equipped with a 12T superconducting allant, Bruker)  - Laser desorption ionization (NdYag 355 nm laser) – Mass spectrometry, at the COBRA laboratory (University of Rouen, France).

The mass spectrometer was externally calibrated with a solution of sodium trifluoacetate. The solid was deposited on an LDI plate (solvent-free), following a procedure previously published by Barrere et al. (2012). Studies have shown that, depending on the ionization mode used, the molecules detected within the analogues show significant differences (Carrasco et al., 2009; Somogyi et al., 2012). Positive ionization is the most widespread in the community for this type of analysis on these analogues and was the only considered in this study. Mass spectra were therefore measured in positive mode at 8 million points with an accumulation of 200 scans, giving a resolution of 600,000 at *m/z* 150 and 100,000 at *m/z* 500. These measured spectra were then calibrated in-house with signals assigned in previous studies, enabling us to obtain a mass accuracy of less than 300 ppb in the mass range considered (Maillard et al., 2018). The instrumental parameters used on similar analogs by Maillard et al. (2018) were applied: Plate offset 100V, deflector plate 210V, laser power 19%, laser shots 40, laser shot frequency 1000 Hz, funnel 1 to 150V, skimmer 1 to 25 V. Peak selection was performed with a signal-to-noise ratio of 5 and 0.01% of intensity. Molecular formulae assigned to each m/z ratio detected with high precision were obtained using the Smartformula tool in Bruker Data Analysis 4.4 software with the following parameters: $C_{0-x}H_{0x}N_{0x-2}$, even and odd electronic configuration allowed, 0.1 ppm error tolerance.

During aerosol harvesting, aerosols are exposed to ambient air. Carrasco et al. (2016) have shown that Titan's analogues produced by the PAMPRE experiment are subjected to oxidation, as with other plasma-produced Titan's organic analogues (McKay, 1996; Tran et al., 2003; Swaraj et al., 2007; Hörst et al., 2018). To limit oxidation of our samples between harvesting and analysis, they are stored and transported in a primary vacuum chamber ($10^{-2}$ mbar). The number of oxygenated molecules detected corresponds between 30% and 50% of the total signals measured. This participation is not negligible, but nevertheless as shown by Maillard et al. (2018) oxygenated species ($O_x$ or $N_xO_x$) are formed from the solid, where no modification was observed between these oxygenated and non-oxygenated species with respect to unsaturation number or H/C ratio. In this study, all oxygenated species are filtered/removed from the data of each Titan aerosol analogues production. Moreover, solid chemical analysis were concentrated on the main contributors by filtering out molecules with an *m/z* ratio greater than 600. The large quantities of data acquired by FTICR-LDI-MS were processed using the "Python tools for complex matrices molecular characterization" (PyC2MC) software by Sueur et al. (2023), enabling to represent graphically the molecular composition of complex materials such as Titan's organic analogues.

## III - Results:

### III.a - Analysis of the gas phase:
### III.a.1 - Initial and final/stable gas composition:

Before starting the experiments, the outer walls of the PAMPRE reactor and the ion-optical section of the mass spectrometer are heated to 120°C, and a plasma of pure $N_2$ is triggered in the plasma cage, to desorb the water present on the inner walls. After 3 hours of this initial heating step, the reactor and spectrometer are pumped in secondary ($10^{-6}$ - $10^{-7}$ mbar) for at least 24 hours. To take account of any residual contamination of the oxygenated ambient air in subsequent analyses, ten measurement cycles in RGA acquisition mode are run on the mass spectrometer in secondary vacuum (named White mass spectrum). After injection of the initial gas mixture of 80% $N_2$ and 20% $CH_4$ stabilized on pressure, and before trigger of the plasma discharge, 10 measurement cycles in RGA acquisition mode are performed to characterize the initial un-ionized $N_2$ - $CH_4$ gas mixture (called OFF mass spectrum). By triggering the plasma discharge, measurements in MID acquisition mode are launched in parallel (described in the next section), then when the intensities tracked for several m/z stabilize and are considered to no longer vary over time (Sciamma'O-Brien et al., 2010), 10 measurement cycles in RGA acquisition mode are carried out to characterize the ionized and stabilized $N_2$ - $CH_4$ gas mixture, i.e. the final/stable gaseous chemical composition set up within the PAMPRE plasma with the experimental parameters set for this study (named ON mass spectrum). The intensities retained for each *m/z* detected on an RGA mass spectrum (White, OFF, ON) are obtained by averaging 10 measurement cycles. For each experimental gas phase analyzed (OFF and ON), the White spectrum (figure 1.A) is subtracted, then the intensities are normalized by the most intense measured at *m/z* 28 corresponding to the presence of $N_2$ (figure 1.B).

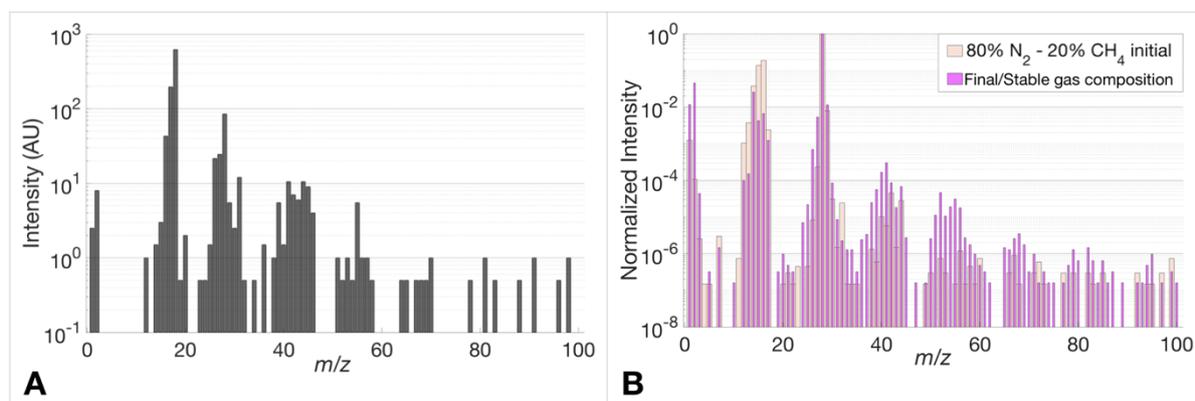

*Figure 1 - Mass spectra acquired on RGA acquisition mode. (A): Averaged mass spectrum from the instrument and the PAMPRE dusty plasma reactor in secondary vacuum. (B): Averaged and normalized mass spectra of experimental gas phases analysed.*

Concerning the White spectrum in figure 1.A, the *m/z* ratios detected show the presence of a water residue (main peak at *m/z* 18), whose abundance appears low, with a measured intensity slightly above the detection limits prudentially considered for the QMS used at $10^2$ counts. The other *m/z* detected are below detection limits and are considered to be the instrument's background noise.

In figure 1.B, the superposition of the OFF (neutral-colored bars) and ON (magenta-colored bars) spectra shows that for *m/z* ratios 16 and 15 corresponding mainly to the presence of $CH_4$ and its fragment $CH_3$ respectively, their measured intensities decrease when the gas composition is stable (ON) compared to the initially injected 20% $CH_4$ (OFF), demonstrating the $CH_4$ consumption achieved when the gas phase varies (detailed in the section III.a.2). The formation of molecular clusters can also be seen on the ON spectrum (figure 1.b), as shown on the analyses of the final/stable gaseous chemical composition by Carrasco et al. (2012) and Dubois et al. (2018) injecting initial proportions of 1 to 10% $CH_4$ into the PAMPRE reactor. These clusters can be defined as Cn families, corresponding to molecular classes with $C_xH_yN_z$ nomenclature for N-bearing species and hydrocarbons, where x+z=n (n is related to the presence of heavy, N-bearing aliphatic compounds). By initially injecting 1 and 10% $CH_4$, Dubois et al. (2018) show that C1 cluster containing $CH_4$ ion peaks and $H_2O$ signatures, and C2 cluster are still present, but whose measured intensities decrease in a manner correlated to the decrease in the initial proportion of $CH_4$. Dubois et al. (2018) also observe that the formation of clusters comprising heavier compounds (C3, C4, C5, and more) is favoured with 10% initial $CH_4$, i.e. when the quantity of $CH_4$ consumed is high as shown by Sciamma'O-Brien et al. (2010).

In this study, by injecting an initial proportion of 20% $CH_4$ into the PAMPRE reactor, the clusters of molecules formed seem similar to those obtained with initial $CH_4$ proportions (1 at 10%) correlating with those observed at different altitudes in Titan's atmosphere. The main *m/z* measured are identical to those identified in advance, in particular the C2 cluster shows strong intensities at *m/z* 26, 27 and 30 which have been attributed to acetylene $C_2H_2$, hydrogen cyanide HCN and ethane $C_2H_6$ respectively (Gautier et al., 2011; Carrasco et al., 2012; Dubois et al., 2018). Clusters C3 and C4 are dominated respectively by *m/z* 41 corresponding to acetonitrile $C_2H_3N$ (Dubois et al., 2018), and by *m/z* 52 attributed to cyanogen $C_2N_2$ as well as a more minority *m/z* at 51 corresponding to cyanoacetylene $HC_3N$ (Gautier et al., 2011; Carrasco et al., 2012). These 6 gaseous products have been identified as potential precursors in the mechanisms of solid formation and growth, both for the aerosol analogues produced by the PAMPRE experiment, but also for Titan's atmospheric aerosols. We find that despite the injection of an initial $CH_4$ proportion of 20% that is not representative of the atmospheric conditions observed in Titan's atmosphere, the gaseous species likely to form and play a part in the satellite's organic chemistry are similar/proximate in our experiments. Thus, the experiments presented here may bring together conditions representative of Titan's atmospheric organic chemistry, but we certainly assume to amplify gaseous kinetic processes (production and consumption), as well as solid growth processes, to facilitate their observations.

What's more, the exposure time of the solids in the plasma is not the same during the different productions of aerosol analogues presented here, i.e. the proportions of gases, notably $CH_4$, are not the same during each production. Indeed, as detailed in the next section, gas kinetics are transien, particularly during the 6 aerosol analogues productions carried out with a pulsed plasma discharge. The proportions of $CH_4$ consumed and remaining during the different durations set to achieve solid productions were determined (summarized in table 2), by multiplying the intensity measured at the time points corresponding to the ON periods set to plasma discharge (MID tracking at *m/z* 16 of $CH_4$, figure 2) by the ratio between the initial 20% of $CH_4$ and the intensity measured at *m/z* 16 on the RGA OFF spectrum (figure 1.B) or that measured at t=0s on the MID tracking of *m/z* 26 (figure 2). For the aerosol analogues Pul20s to Pul100s, we observe that, depending on the duration of the discharge, the proportion of $CH_4$ consumed or the proportion of $CH_4$ remaining can be correlated with Titan proportions of 1 to 10% (table 2). The question is which proportion of $CH_4$ (initial or consumed or remaining) is the most significant to consider in representing the percentage of $CH_4$ used by a laboratory experiment to simulate Titan's organic chemistry.

**Table 2 -** Proportions of $CH_4$ consumed and remaining calculated in relation to the initial $CH_4$ proportion of 20%, at different durations during the experimental simulation set up with the parameters fixed at the PAMPRE reactor in this study.

| Plasma discharge time (s) | Name of aerosol analogues production | % $CH_4$ consumed | % $CH_4$ remaining |
| --- | --- | --- | --- |
| 20 | Pul20s | 5.54 ± 0.50 | 14.46 ± 0.50 |
| 35 | Pul35s | 8.94 ± 0.63 | 11.06 ± 0.63 |
| 50 | Pul50s | 12.84 ± 0.71 | 7.16 ± 0.71 |
| 75 | Pul75s | 17.18 ± 0.79 | 2.83 ± 0.79 |
| 100 | Pul100s | 18.91 ± 0.77 | 1.09 ± 0.77 |
| 145 | Pul145s | 19.42 ± 0.66 | 0.58 ± 0.66 |
| continue | Continuous | 19.41 ± 0.62 | 0.59 ± 0.62 |

## III.a.2 - Kinetic evolution of methane and some gaseous products:

Using the PAMPRE dusty plasma reactor, previous studies have shown that certain experimental parameters impact $CH_4$ dissociation kinetics. With a higher injection rate (55 sccm), Sciamma O'-Brien et al. (2010) showed that: $CH_4$ dissociation time extends from 28s to 1min40s, when increasing the initial proportion of $CH_4$ from 1 to 10%; $CH_4$ consumption efficiency increases when decreasing its initial proportion, and leads to a variation in the consumed and stable/constant amounts of $CH_4$. Wattieaux et al. (2015) also observed that, in parallel with $CH_4$ consumption (transient kinetic phase), electron density decreases simultaneously with an increase in electron temperature, leading to accelerated $CH_4$ consumption before stabilization in the gas phase.

In this study, after injecting and stabilizing at pressure the initial gas mixture of $N_2$ - $CH_4$, we begin time-tracking measurements of several m/z ratios corresponding to different targeted neutral species (discharge OFF). After one minute of acquisition, we trigger the plasma discharge while continuing to follow the temporal evolution of these masses until the intensities measured are constant over time. By decreasing the injection rate to 2.5 sccm and increasing the initial proportion of $CH_4$ to 20%, we observe on the $CH_4$ time tracker shown in figure 2 that the $CH_4$ dissociation time is extended to 2min30s (figure 2; table 3). As observed by Sciamma O'-Brien et al. (2010) and Wattieaux et al. (2015), the MID tracking of $CH_4$ represented in figure 2 shows two main kinetic phases: a first phase where $CH_4$ is consumed (transient), and a second where it reaches a constant concentration over time (stable). At the end of the transient regime, the proportion of $CH_4$ consumed is around 19%, leaving a constant $CH_4$ concentration at steady state of less than 1% in he gas mixture. As observed by Wattieaux et al. (2015), the transient kinetic phase of $CH_4$ consumption is not constant (figure 2), and presents an inflection point dissociating two kinetic sub-phases characterized by a different experimental reaction rate constant called $k_{1,CH4}$ and $k_{2,CH4}$.

The first consumption sub-phase follows a decreasing exponential function, allowing the experimental reaction rate constant $k_{1,CH4}$ ($s^{-1}$) to be characterized by relationship (1) :

$$I = (I_0 - I_{stable}) \times e^{-k_{1,CH4} t} + I_{stable} \quad (1)$$

where I (U.A) is the intensity measured on MID monitoring for a gaseous species. $I_0$ is the initial measured intensity of the species (before discharge initiation). $I_{stable}$ is the intensity measured during the stable phase extrapolated for this first phase of methane consumption before the inflection point. t is the time (s).

The decreasing exponential represented on the $CH_4$ monitoring (yellow curve in figure 2) shows that, after a certain time, the mean constant of $k_{1,CH4}$ (table 2) set is no longer representative of the measurement points (appearance of an inflection point), distinguishing a second kinetic sub-phase of $CH_4$ consumption. From this inflection point, the rate constant $k_{2,CH4}$ ($s^{-1}$) is calculated according to relationship (2):

$$I = (I_{flexion} - I_{stable}) \times e^{-k_{2,CH4}(t-t_{flexion})} + I_{stable} \quad (2)$$

where $t_{flexion}$ corresponds to the time (s) at which the inflection point is positioned, depending on the gas species. $I_{flexion}$ corresponds to the intensity measured at this point.

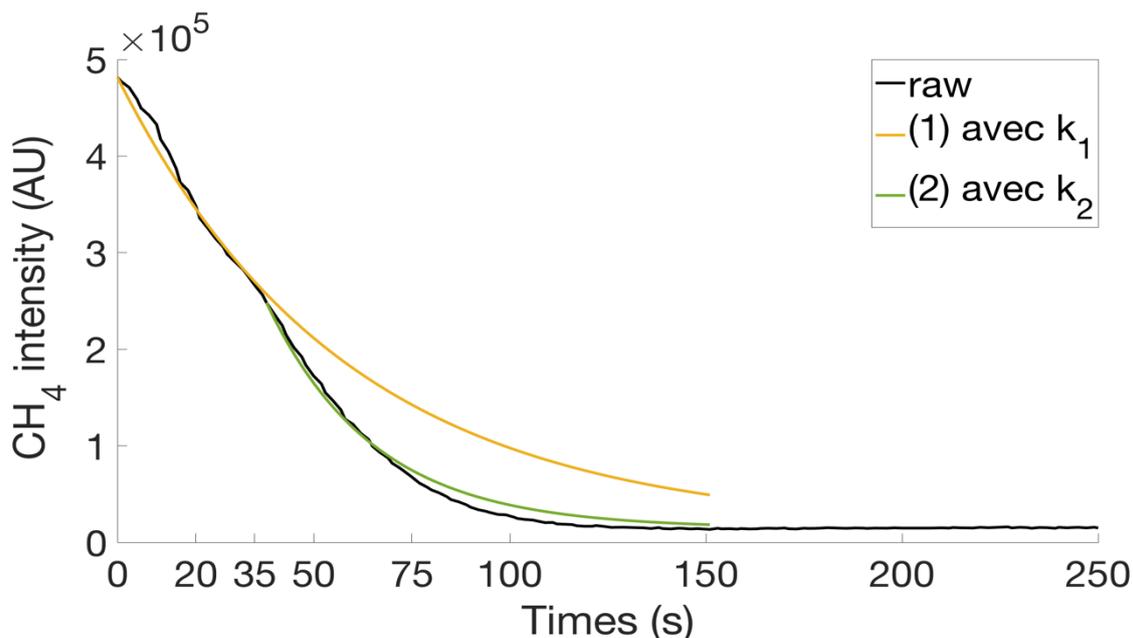

*Figure 2 - Time MID tracking of intensity of m/z 16 attributed to the presence of $CH_4$. At t=0s, the plasma discharge is triggered within the experimental simulation. The black curve tracks the raw intensities measured. The yellow and green curves correspond to the intensities calculated from relations (1) and (2) by fixing the calculated mean values of the experimental reaction rate constants $k_{1,CH4}$ and $k_{2,CH4}$ (table 2).*

Previous studies have shown that $CH_4$ consumption kinetics influences the overall chemical composition when the gas phase stabilizes. Indeed, Sciamma O'-Brien et al (2010); Gautier et al (2011); Carrasco et al (2012); Dubois et al (2018) have shown that $CH_4$ dissociation impacts the proportions of gaseous products formed, including in particular the predominance ratio between gaseous hydrocarbons and nitrogen-containing products represented mainly by nitriles. However, their observations show the regular presence of certain neutral gaseous products within the plasma even when varying the experimental parameters, and which we also detect with the experimental conditions set in this study. We therefore focused our kinetic analyses by QMS, following the intensities corresponding to the majority peak/fragment identified to characterize the particular presence of following neutral products by mass spectrometry: $C_2H_2$ (*m/z* 26); HCN (*m/z* 27); $C_2H_6$ (*m/z* 30); $C_2H_3N$ (*m/z* 41); $HC_3N$ (*m/z* 51) and $C_2N_2$ (*m/z* 52).

At the *m/z* ratio considered principal for $C_2H_2$ (26), we also take into account the presence of another gaseous species, namely HCN. In electron ionization mass spectrometry, a given gaseous species presents a main peak at a fixed *m/z* ratio, but also other minority peaks at different *m/z*, which are due to the fragmentation of the gaseous species within the QMS ionization chamber (Gautier et al., 2020). For a given gas species, the characteristic *m/z* ratios (main and minority) are always the same, and the intensities of the minority peaks each represent a certain proportion of that of the main peak. According to the NIST database, HCN has a contribution at *m/z* 26, representing around 17% of its main intensity set at *m/z* 27. From the intensities measured over time (time tracking) with an *m/z* set at 26, we have subtracted the possible contribution of HCN, removing 17% of the intensities measured on time tracking at *m/z* 27, to observe only the kinetics of the $C_2H_2$ gas species on time tracking at *m/z* 26.

As can be seen from the individual time MID tracking of each neutral gas product shown on figure 3, these appear simultaneously from the start of $CH_4$ consumption (at t=0s), then their intensities increase in correlation with the decrease in $CH_4$ consumption, presenting an initial transient gas kinetic sub-phase referred to here as "gas product production" for simplicity's sake. Indeed, we note that despite the observation of an increase in the intensities of these products demonstrating a certain production, the gases can interact in parallel with their formation and be consumed in gas-phase reactions, for example. At the end of this kinetic sub-phase of production, neutral gases show the highest measured intensities on their time MID tracking (and compared RGA spectra). The experimental reaction rate constants $k_{Prod}$ ($s^{-1}$) characterizing the increasing exponential of gaseous product production kinetics are given by relation (3):

$$I = I_{flexion} \times (1 - e^{-k_{Prod} t}) \quad (3)$$

Still during the transient kinetic phase and starting from a duration depending on the gaseous species, these present an inflection point in their temporal tracking, where the intensities measured for the gaseous products decrease in anti-correlation with that of $CH_4$, eventually stabilizing at an identical duration for all the gaseous species observed. This second transient kinetic sub-phase drives the gaseous products into a net consumption that can be characterized by a rate constant $k_{Cons}$ ($s^{-1}$) following the same decreasing exponential as that described for $k_{2,CH4}$ of $CH_4$ by relationship (2).

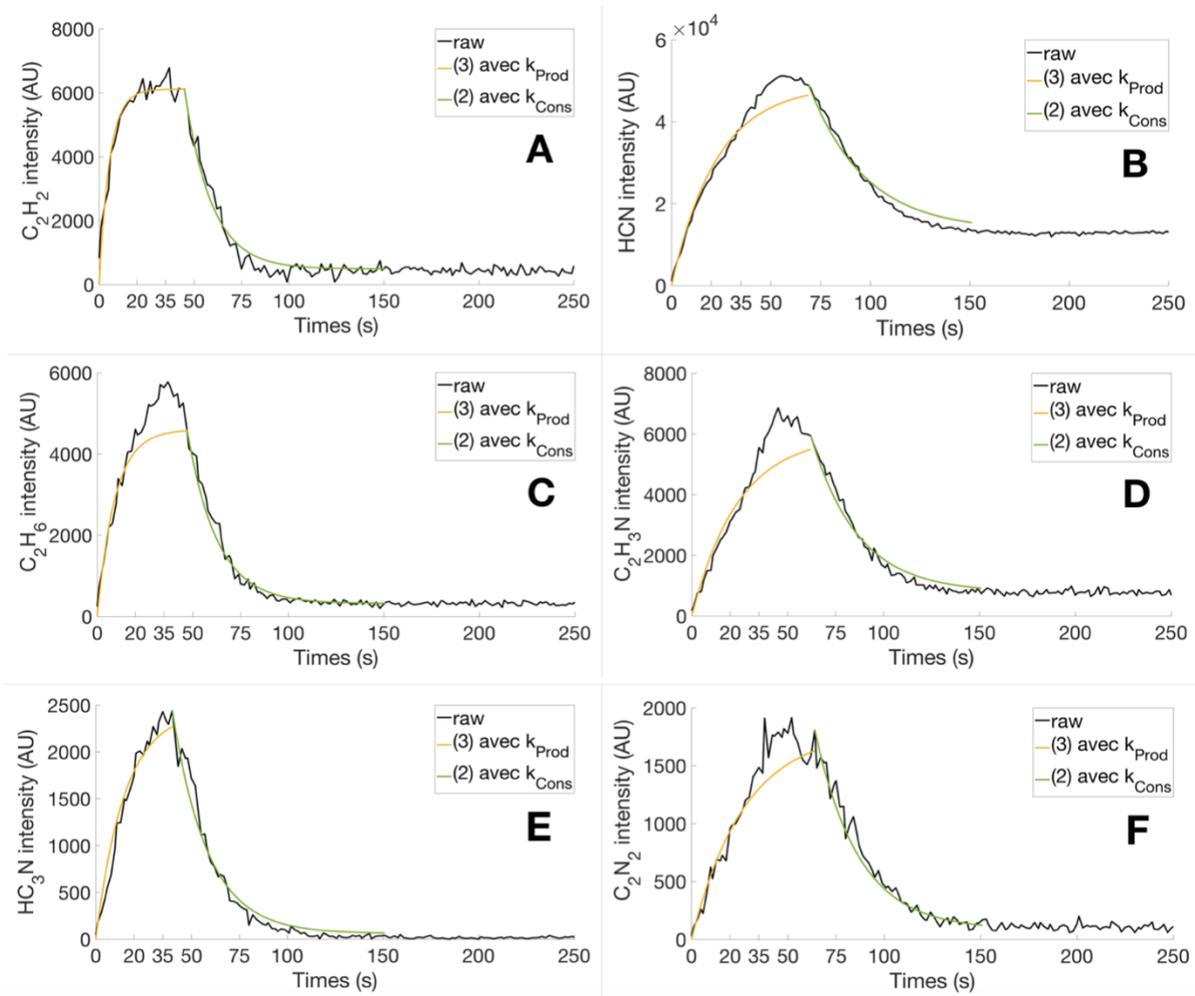

*Figure 3 - Time MID tracking of intensities corresponding to the main m/z considered for the gaseous products studied. At t=0s, the plasma discharge is triggered within the experimental simulation. The black curve tracks the raw intensities measured. The yellow and green curves correspond to the intensities calculated from relations (3) and (2) by fixing the calculated mean values of the experimental*

reaction rate constants $k_{Prod}$ and $k_{Cons}$ (table 2). (A): m/z 26 for $C_2H_2$. (B): m/z 27 for HCN. (C): m/z 30 for $C_2H_6$. (D): m/z 41 for $C_2H_3N$. (E): m/z 51 for $HC_3N$. (F): m/z 52 for $C_2N_2$.

The mean experimental reaction rate constants determined from time monitoring, before the inflection point ($k_{1,CH4}$, $k_{Prod}$) and after it ($k_{2,CH4}$, $k_{Cons}$) are summarized in table 3 for $CH_4$ and the following gaseous products: $C_2H_2$, HCN, $C_2H_6$, $C_2H_3N$, $HC_3N$ and $C_2N_2$. The intensity curves calculated using exponential functions and fixing the mean value considered for the rate constants k (table 3) are shown in figures 2 and 3. Based on the intensities measured on time MID tracking of the various gaseous species, the characteristic durations of the different kinetic sub-phases (consumption and production) have also been recapitulated in table 3. Consumption of gaseous products is quantified, by the drop in intensity measured between the production peak where the proportion of gaseous product is considered maximum in the simulation (equivalent to 100%) and when it becomes constant.

**Table 3 -** Kinetic parameters characterizing the temporal evolution of intensities measured for certain gaseous products and $CH_4$, by in-situ mass spectrometry. $k_{1,CH4}$ and $k_{Prod}$ correspond to the average constants calculated for the experimental reaction rate from relationships (1) and (3) respectively, characterizing the temporal kinetics of gases between the initiation of the plasma discharge (t=0) and the inflection point present on the time tracks. $k_{2,CH4}$ and $k_{Cons}$ correspond to the average reaction rate constants calculated experimentally from relation (2), characterizing the temporal kinetics of the gases after the inflection point and until stabilization of the gas chemistry. The uncertainties of the experimental rate constants are given by the quadratic sum of the uncertainties associated with the measurements used for the calculation, taking the standard deviation of the mean of the measured QMS intensities included in each kinetic sub-phase (QMS measurement fluctuation) and an accuracy of 0.2s for the temporal QMS measurement.

| Transient kinetic regime of gaseous products measured by in-situ mass spectrometry | | | | | | |
|---|---|---|---|---|---|---|
| | | **Production kinetics sub-phase** | | **Consumption kinetics sub-phase** | | |
| Gaseous species | m/z tracking | Time duration (± 0.2 s) | $k_{Prod}$ (±).$10^{-2}$ ($s^{-1}$) | Time duration (± 0.2 s) | $k_{Cons}$ (±).$10^{-2}$ ($s^{-1}$) | % consumed of each gaseous product |
| $C_2H_2$ | 26 | 45 | 18.32 ± 4.50 | 105 | 7.08 ± 1.19 | 92.03 ± 2.87 |
| $C_2H_6$ | 30 | 47 | 10.76 ± 2.78 | 103 | 6.40 ± 1.29 | 93.04 ± 3.25 |
| HCN | 27 | 69 | 4.39 ± 0.13 | 81 | 3.54 ± 0.96 | 72.60 ± 3.25 |
| $C_2H_3N$ | 41 | 62 | 4.14 ± 1.10 | 88 | 4.10 ± 0.85 | 86.80 ± 3.22 |
| $HC_3N$ | 51 | 42 | 5.71 ± 1.42 | 108 | 5.67 ± 1.00 | 96.90 ± 2.58 |
| $C_2N_2$ | 52 | 64 | 3.64 ± 0.80 | 86 | 4.40 ± 1.95 | 95.30 ± 3.40 |
| **Transient kinetic regime of $CH_4$ gas measured by in-situ mass spectrometry** | | | | | | |
| | | **1st sub-phase consumption kinetics** | | **2nd sub-phase consumption kinetics** | | |
| m/z tracking | | Time duration (± 0.2 s) | $k_{1,CH4}$ (±).$10^{-2}$ ($s^{-1}$) | Time duration (± 0.2 s) | $k_{2,CH4}$ (±).$10^{-2}$ ($s^{-1}$) | |
| 16 | | 38 | 1.73 ± 0.48 | 112 | 3.66 ± 0.66 | |
| 15 | | 38 | 1.17 ± 0.10 | 112 | 3.27 ± 0.93 | |

The kinetics of $C_2H_2$, $C_2H_6$ and $HC_3N$ appear to correlate temporally, with a drop in measured intensities (consumption kinetic sub-phase) beginning between 40 and 50 seconds after ignition of the plasma discharge (table 3). We also note that the inflection point observed on $CH_4$ monitoring is positioned at a similar temporal duration (around 38s), where an acceleration of $CH_4$ consumption occurs with $k_{2,CH4} > k_{1,CH4}$ (table 2). For HCN, $C_2H_3N$ and $C_2N_2$, their kinetics appear to correlate in time, where their measured intensities increase (production) up to 60 - 70 seconds (table 3).

The production rate constants $k_{Prod}$ of the two neutral hydrocarbons $C_2H_2$ and $C_2H_6$ are the highest of the six gaseous products measured (table 3). Nitriles such as HCN, $C_2H_3N$, $HC_3N$ and $C_2N_2$ have similar production rate constants, one order of magnitude lower than the hydrocarbons (table 3). The consumption slopes of neutral hydrocarbons ($C_2H_2$ and $C_2H_6$) are characterized by an average rate constant $k_{Cons}$ lower than that of their production ($k_{Prod} > k_{Cons}$), with values particularly close to that of $HC_3N$. As for nitriles such as HCN, $C_2H_3N$, $HC_3N$ and $C_2N_2$, the values of constants characterizing their production and consumption have fairly similar ($k_{Prod} \sim k_{Cons}$).

Although the hydrocarbons show a production slope characterized by a high $k_{Prod}$ constant, the duration of their consumption is almost twice that of their production, resulting in a high consumption of $C_2H_2$ and $C_2H_6$ of around 90% before their concentrations stabilize over time. In the case of nitriles, production and consumption times can vary, particularly between $HC_3N$ and the other products HCN, $C_2H_3N$ and $C_2N_2$. It can be seen that the lighter nitriles (HCN and $C_2H_3N$) are the least consumed products, compared with the heavier nitriles ($HC_3N$ and $C_2N_2$), whose proportions consumed are similar to those of hydrocarbons (around 90 - 95%). We note that these differences in proportions consumed between nitriles may be due to the maximum available proportion of the gaseous product after the sub-phase of production, which is not the same depending on the species considered (higher for lighter species).

During the kinetic sub-phase of production of the various gaseous products (figure 3), the slope characterizing the increase in measured intensities attenuates and begins to form a plateau (before peak production). This attenuation suggests that another process consuming the gaseous products is interacting in parallel with their production and becoming increasingly dominant, until perhaps converting the kinetic production of the gaseous products into a net consumption. However, the influence and interaction with this new consuming factor looks different depending on the gaseous product considered, in view of the kinetic variations (temporal, rate constant k) observed particularly between gaseous hydrocarbons and nitrogenous species. Wattieaux et al. 2015 showed that the acceleration-inducing electron temperature rise during $CH_4$ consumption was due to the appearance of the first solid particles within the plasma, as previously observed by Alcouffe et al. (2010). An initial study of the conversion of gas to solid particles in the PAMPRE experiment by Sciamma'O-Brien et al. (2010) quantified the mass production rate to solid aerosols as a function of the rate of conversion to carbon from $CH_4$ fragmentation/dissociation. They showed that the rate of solid aerosol production can be represented by a parabolic function, highlighting two competitive chemical regimes controlling differently the efficiency of the aerosol production. When nitrogen-containing molecules dominate the gaseous chemical composition, the mechanisms tend to favor the growth of solids rather than their production, and conversely the predominant presence of gaseous hydrocarbons favors the production of solids rather than their growth (Sciamma'O-Brien et al., 2010; Carrasco et al., 2012).

The experimental reaction rate constants determined for some gaseous products and $CH_4$ may represent interaction rates involving gases in gaseous chemical reactions leading to their production or consumption, but also with a close order of magnitude (around $10^{-2}$ $s^{-1}$) these may testify to the influence of a common process impacting the kinetics of gaseous species and involving the appearance of solid aerosols carrying out gas-consuming interactions. As highlighted for HCN in the phenomenological study presented in Perrin et al. (2021), we can assume that these consumption reactions are linked to heterogeneous processes between gaseous compounds and the solid particles being formed/growth. In this case, the calculated rate constants are directly dependent on several quantities: 1/ an intrinsic characteristic of the interaction between the gas and the particle: the sticking coefficient, and 2/ extrinsic characteristics of this interaction linked in particular to the size and average concentration of particles within the plasma. The results of the morphology and chemical composition analyses carried out on the analogous aerosols produced here will enable us to characterize these extrinsic properties specific to the experimental conditions chosen, and then to deduce the bonding coefficient of potential reactions between gases and solids.

### III.b - Morphological analysis of Titan aerosol analogues

Like Tomasko et al. (2005; 2008), aerosol analogues as individual entities will be referred to as monomers (otherwise named grains, particles by other studies). Using the PAMPRE experiment, studies producing different Titan analogues observed monomers with a generally quasi-spherical morphology and a rough surface (Hadamcik et al., 2009). Varying certain experimental parameters impacted the size distribution, forming monomers with diameters ranging from 0.01 - 6 µm. With a continuously operating RF discharge, it has been observed that the formation of solid monomers takes place with minimum gas concentrations (Dubois et al., 2018) and that the expulsion of monomers out of the plasma (the end of monomer growth) certainly takes place when gas concentrations are constant/stable (Sciamma'O-Brien et al., 2010).

### III.b.1 - Microphysical evolution of aerosol analogues according to the duration of the plasma discharge:

Figure 4 shows the SEM images obtained of Titan aerosol analogues as a function of the time at which the plasma discharge was switched on (table 1). The aerosols produced using the experimental conditions set at the PAMPRE experiment in this study are also predominantly in the form of quasi-spherical monomers, with sizes ranging from nanometric to micrometric. With the shortest residence time set for monomers within the plasma (20s), the production called Pul20s is considered to represent the first generation of monomers formed, i.e. the physico-chemical properties determined on this aerosol population are considered those initially produced/formed using the experimental conditions set at the PAMPRE reactor in this study. When the discharge time exceeds 20 seconds, the solid monomers produced show different morphology (this section) and chemical composition (section III.c) from those initially observed on Pul20s aerosol analogues.

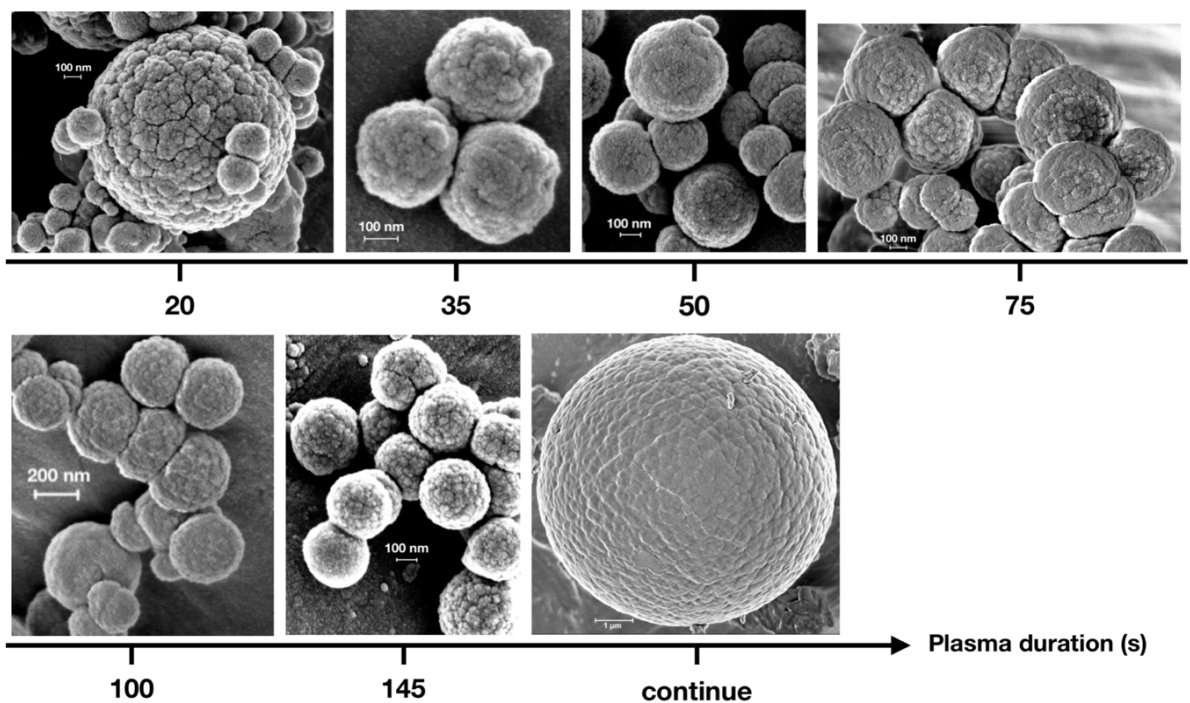

*Figure 4 - SEM images of Titan aerosol analogues produced with different residence/exposition times in the plasma of the PAMPRE experiment by setting the other experimental parameters (Pul20s, Pul35s, Pul50s, Pul75s, Pul100s, Pul145s and Continous from table 1).*

The diameter distributions are determined from the processing of these SEM images, and are summarized in table 4. The first generation of monomers (Pul20s) has the narrowest range of measured diameters, with the smallest mean diameter (around 200 nm). However, Pul20s production presents monomers with a bimodal size distribution, including one with a larger diameter of the micrometer-size order (table 4). This micrometer size population is discussed in section III.b.2. Increasing the duration of the plasma discharge produces aerosols with a wider diameter range from 249 to 766 nm (Pul35s and Pul145s respectively), although the average diameters measured remain very close between 350 and 500 nm for Pul35s, Pul50s, Pul75s, Pul100s and Pul145s. As for monomers produced with an unknown residence time in the plasma (Continuous production), the range of diameters measured increases sharply to a mean diameter retained multi-micrometer (table 4).

**Table 4 -** Range of diameters measured on SEM images of solid monomers produced with different residence times in the plasma of the PAMPRE experiment. Uncertainties on the mean diameters used are given by the standard deviation of the mean of all diameter measurements made for each analogous aerosol production.
*Minimum and maximum diameters
** determined from 100 spherical monomers for each production in pulsed mode and 32 monomers for Continuous production. The monomers counted were detected on different SEM images (at least 10), each taken on a different area of the grid surface where aerosol analogues were deposited.

| Experimental Condition | | SEM Analysis | |
|---|---|---|---|
| Plasma discharge time (s) | Production names | Measured diameter range** (nm) | Average diameter** (nm) |

| | | 131 - 291 | 193 ± 6 |
| --- | --- | --- | --- |
| 20 | Pul20s | 803 - 1460 | 1190 ± 294 |
| 35 | Pul35s | 249 - 454 | 347 ± 13 |
| 50 | Pul50s | 275 - 531 | 397 ± 14 |
| 75 | Pul75s | 304 - 759 | 483 ± 19 |
| 100 | Pul100s | 292 - 564 | 404 ± 17 |
| 145 | Pul145s | 309 - 766 | 509 ± 24 |
| continue | Continuous | 5304 - 16680 | 8787 ± 794 |

The first generation of monomers (Pul20s) appears in parallel with the kinetic sub-phase of gaseous product production, and before the acceleration of the $CH_4$ consumption rate in the gas phase. The diameters measured on the productions made with a more long controlled plasma discharge remain well below the micrometric monomers produced with a continuous plasma discharge, which means that the monomers of the Continuous production have probably resided within the plasma for longer than 145 - 150 seconds and have finished their growth when the gas concentrations are constant/stable. As shown by Sciamma'O-Brien et al. (2010) and Carrasco et al. (2012), depending on the chemical nature of the gaseous species predominating the composition of the gas phase, the chemical mechanisms occurring during our different aerosol productions may rather favor the processes of solid production than those of solid growth, or vice versa.

To get an idea of the microphysical evolution observed on Titan analog aerosols (Productions Pul), we will characterize the spherical growth in size/volume of monomers by the temporal evolution of the diameter range measured of the first generation of monomers (Pul20s) to on more advanced aerosols produced with a longer plasma discharge duration (figure 5.A). To characterize the rate of solid aerosol production as a function of RF plasma discharge duration, we estimate the average number of solid monomers present in the plasma. For this purpose, each individual monomer collected is considered to have remained for the assigned duration in the plasma discharge called $t_{ON}$. The total time of the discharge was actually on $t_{Tot,ON}$ (s, Eq. 4) is considered to be that which produced the total harvested mass of aerosols $m_{Tot,harvested}$ (g, table 1), over the total production time $t_{Tot,Prod}$ (s, table 1) taking into account discharge off times. The total number of aerosols harvested $N_{Part,tot}$ and the average number of aerosols present in the plasma $N_{Part,moy}$ are obtained by relations (5) and (6) respectively, where the volume of a monomer $V_{aerosol}$ ($cm^3$) is calculated by considering the average diameters measured by SEM on matching spherical monomers (cm, table 4).

$$t_{Tot,ON} = (\frac{t_{Tot,Prod}}{(t_{ON} + 600)}) \times t_{ON} \quad (4)$$

$$N_{Part,tot} = \frac{m_{Tot,harvested}}{(\rho_{aérosol} \times V_{aérosol})} \quad (5)$$

$$N_{Part,moy} = \frac{N_{Part,tot} \times t_{ON}}{t_{Tot,ON}} \quad (6)$$

where $\frac{t_{Tot,Prod}}{(t_{ON} + 600)}$ corresponds to the number of cycles (ON and OFF periods) completed during the total duration of an aerosol production $t_{Tot,Prod}$. The constant 600 corresponds to the number of seconds the plasma remains off between each ON period (constant for all productions and equal to 10 min).

The density $\rho_{aerosol}$ of the various aerosol analogues produced in this study is not known. Other experimental studies producing Titan solid analogues using plasma experiments, as well as those produced by the PAMPRE reactor, have calculated their densities using different techniques and are summarized in table 5 (Trainer et al., 2006; Imanaka et al., 2012; Hörst et al., 2013; Brouet et al 2016). These vary between 0.4 and 1.44 g.cm$^{-3}$ depending on the experimental formation condition (initial $CH_4$ proportion, pressure within the plasma). For the rest of the calculations, we use these density values for our aerosol analogues.

**Table 5** - Effective density values determined on Titan solid analogues formed using plasma experiments by Hörst and Tolbert (2013) and Imanaka et al. (2012), and the PAMPRE experiment by Brouet al. (2016).

| Studies | % initial $CH_4$ | Average monomer diameters (nm) | Effective density (g.cm$^{-3}$) |
|---|---|---|---|
| Hörst et Tolbert, 2013 | 2 | 42.3 ± 1.9 | 1.13 ± 0.1 |
| | 5 | 20.0 ± 1.2 | 0.66 ± 0.1 |
| | 10 | 14.4 ± 0.6 | 0.4 ± 0.1 |
| Imanaka et al., 2012 | 10 | - | 1.3 – 1.4 |
| Brouet et al., 2016 | 5 | 400 - 500 | 1.44 ± 0.01 |

For each aerosol analogues production (volume of an individual monomer known), the $N_{Part,tot}$ was calculated with relation (4) for different density values determined in the literature. Shown in figure 5.B, the total number of particles collected decreases with increasing density, but the variation is not even an order of magnitude between the smallest and largest density values (0.4 and 1.44 g.cm$^{-3}$). For the $N_{Part,moy}$ calculation, a density value of 0.4 g.cm$^{-3}$ and 1.44 g.cm$^{-3}$ was used in relation (5) for all aerosol analogues productions. Indeed, the value of 1.44 g.cm$^{-3}$ corresponds to that determined by Brouet et al. (2016) on aerosol analogues produced using the PAMPRE experiment, but with an initial $CH_4$ proportion of 5% and a higher pressure than that set here. In this study, by increasing the initial proportion of $CH_4$ to 20% (with lower pressure), the density of the aerosol analogues may be different from Brouet et al. (2016), considering that increasing the initial proportion of $CH_4$ seems to lead to a decrease in the density of the solids produced (table 5) (Hörst and Tolbert (2013)). The smallest measured density value (0.4 g.cm$^{-3}$) is therefore also taken into account, extracted from Hörst and Tolbert (2013). For each fixed plasma discharge duration, $N_{Part,moy}$ is plotted in Figure 5.C. The uncertainty of $N_{Part,moy}$ is given by a quadratic sum, taking into account the standard deviation of the mean of $N_{Part,tot}$ values calculated with the different density values. The uncertainty deduced here for $N_{Part,moy}$ gives a minimum value for it, taking no account of aerosol analogues losses during harvesting, which are impossible to estimate.

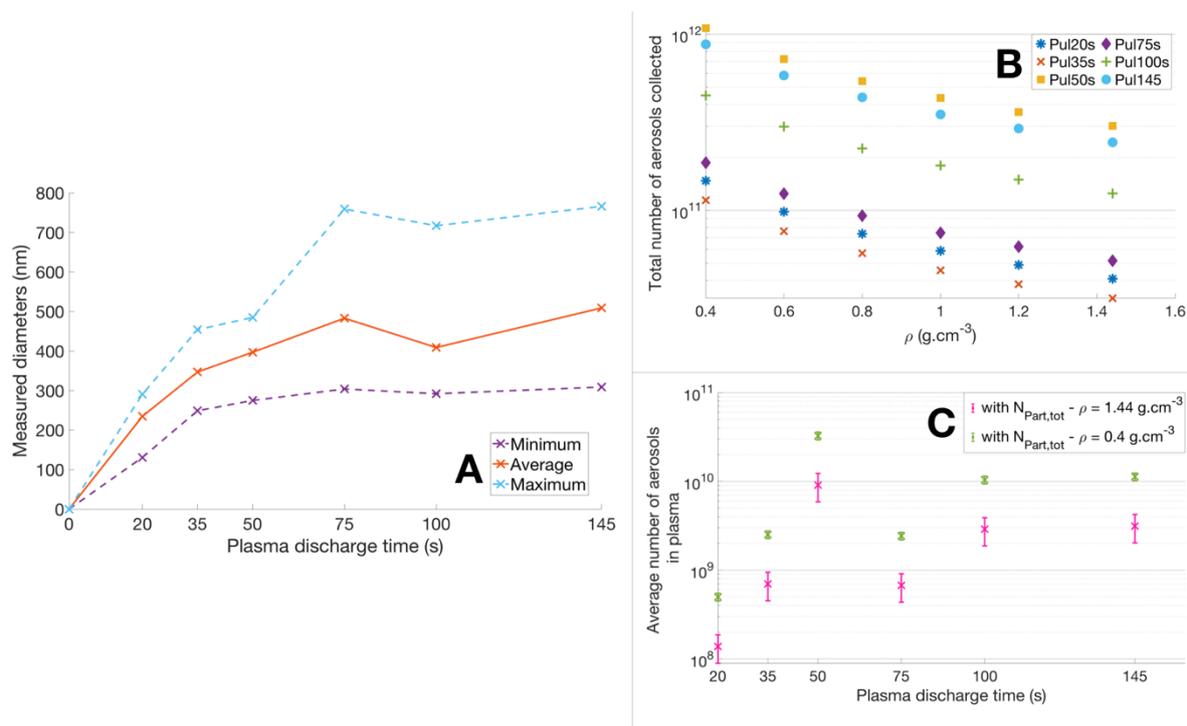

*Figure 5 - (A): Time evolution of spherical volume growth of Titan aerosol analogues produced with different residence/exposition times within the plasma. (B): Total number of analog aerosols harvested $N_{Part,tot}$ determined by relation (4) as a function of density variation, for each aerosol analogues production. (C): Average number of solid aerosols/monomeres present in the plasma $N_{Part,moy}$ determined by relation (5) as a function plasma discharge duration. The density was set at 0.4 and 1.44 g.cm$^{-3}$ with $N_{Part,tot}$ concordant in agreement with the extreme values found in the literature (Hörst and Tolbert 2013 and Brouet et al. 2026).*

Up to a time of around 50 seconds, the monomers produced show volume growth with a doubling range of nanometric diameters (maximum, average, minimum; figure 5.A). In parallel, the calculated average number of solid monomers present in the plasma increases by almost two orders of magnitude (time < 50s in figure 5.C, for each value density fixed). From 50 to 145 seconds, the spherical growth in volume of the monomers produced continues to expand slightly, yet the average aerosol morphology of the Pul50s, Pul75s, Pul100s and Pul145s productions shows a similar diameter around 400 - 500 nm. At the same time, the average number of monomers calculated decreases from 50 to 75 seconds, then $N_{Part,moy}$ increases again, but with a less steep slope after 75 seconds. The evolution of the volume growth and production of solid monomers seems to dissociate different temporal periods favoring their formation $^{and}/_{or}$ growth and correlating with the observed kinetic temporal evolution of gaseous products. These correlations are discussed in detail in the following, involving in particular the change in the nitrogen chemistry observed in the gas phase.

Over a period of time from plasma initiation (t=0s) to 45-50 seconds, gaseous products, particularly gaseous hydrocarbons, exhibit a kinetic sub-phase of production where their abundances increase within the gas phase (figure 2), and whose presence may favor the production of new solid aerosol nuclei rather than the growth of nuclei already formed, as shown by Sciamma O'-Brien et al. (2010) and Carrasco et al. (2012). From 35s onwards, the monomers also show the start of volume growth, which remains fairly restricted as gas concentrations vary, then becomes very significant when concentrations are constant over time. According to the observations made here and correlating with those made in previous studies, the growth mechanisms of the solids are certainly influenced more by the nitrogenous gaseous species. Indeed, the calculated average experimental reaction rate constants vary very slightly over time for nitriles (table 2), which could be interpreted as a constant participation of these nitrogenous products along the aerosol analogues growth that starts early in our experiments. A change in nitrogen chemistry is also observed in the analyzed composition of aerosol analogues produced with different growth times (discussed in the next section). Furthermore, the very significant volume increase observed on resident monomers over an unknown period of time when gas concentrations are constant, occurs after a significant amount of gaseous products have been consumed during the transient kinetic phase, notably the nitriles such HCN.

Over time, the number of solid monomers present in the plasma stabilizes around a maximum calculated at $10^{+10}$ with a tendency to decrease, when the gaseous products enter their kinetic sub-phase of consumption. At the same time, particular morphologies appear where several monomers coagulating with each other, which can lead to a reduction in the number of monomers. Beyond a certain number of monomers present in the plasma (estimated at around $10^{+10}$), the probability of collision of solid monomers certainly increases, and thus favors the formation of the morphologies described in the following section, demonstrating possible mechanisms of solid growth controlled by microphysical laws, as showed Lavvas et al. (2011).

### III.b.2 - Particular morphologies observed in Titan aerosol analogues:

In the studies producing Titan aerosol analogues, certain morphologies have already been observed. Using the PAMPRE reactor, Hadamcik et al. (2009) found that micrometer-sized spherical monomers appeared to represent coarse, brittle aggregates composed of smaller elements probably oriented during their growth by electrostatic charges providing them with a radial structure. Sciamma O'-Brien et al. (2017) also observed different morphologies on Titan aerosol analogues produced using a cold plasma experiment (Sciamma'O-Brien et al., 2014), where this exhibit three main classes of microphysical structures: nanometric particles with dimensions between 10 and 50 nm, quasi-spherical grains with diameters measured between 100 and 500 nm, and several aggregated grains (aggregate) reaching dimensions between 1 and 5 µm. They suspect that the grains measuring several hundred nanometers are made up of a cluster of nanometric particles, some fragments of which have been observed individually.

A particular morphology appeared only for the Pul20s aerosol analogues production, is shown in figure 6. The diameter distribution measured on the SEM images was split into two, as shown in table 4. A first distribution of nanometric diameters lies at around 193 nm, and a second is positioned at micrometric upper mean diameters of around 1 µm. Some solid monomers must have sedimented on the edges of the plasma cage, then resuspended an unknown number of times during the numerous repetitions of the ON period during production, allowing their growth to continue. These SEM images (figure 6) reveal spherical nanometric particles embedded directly on the surface of spherical micrometric particles. The first generation of nanometric monomers can be attracted at the surface of larger micrometric monomers, who have a effective collision cross-section increasing. Up to a discharge duration of 20s, gaseous products are increasingly abundant (kinetic sub-phase of production), certainly contributing to the realization of a surface chemistry allowing the coagulation of primary monomers until reforming a new, more voluminous monomer retaining a quasi-spherical shape, as described by Lavvas et al. (2011) or observed by Hadamcik et al. (2009) and Sciamma O'-Brien et al. (2017).

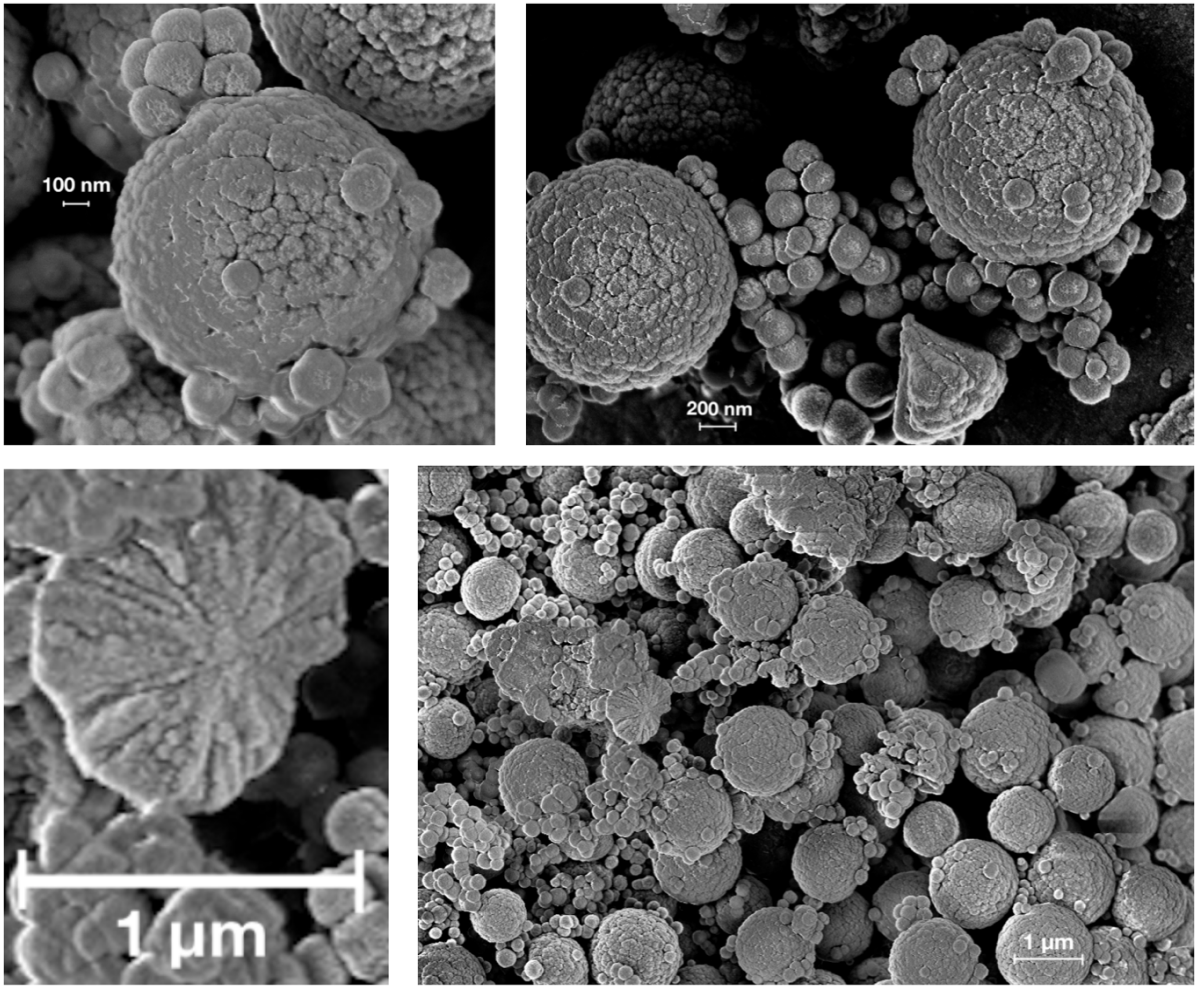

*Figure 6 - Particular morphologies of spherical aggregate structures formed by the coagulation of spherical monomers, observed within Titan Pul20s analog aerosol production.*

      Other particular and repetitive morphologies made up of solid monomers are observed on the SEM images on figure 7. With a residence time equivalent to or greater than 50 seconds, microphysical structures of aggregates comprising several individual monomers coagulating together to form a single spherical monomer of larger size, appear on images A, B and C in figure 7. This mechanism is similar to that observed for Pul20s production (figure 6), but the aerosols coagulating with each other are larger (around 350 - 500 nm) and seem less numerous, compared to Pul20s monomers (figure 6). This phenomenon probably occurs more slowly than the chemical reactions over time and may explain the slight increase in aerosol volume growth observed on figure 5.A. Only in the case of production using a continuously operating plasma discharge is the formation of aggregates made up of more than a dozen individual monomers observed in images E, F and G of figure 7. The individual monomers in these aggregates have average diameters of several µm. Within the aggregates themselves, monomers appear to be coagulating with each other (image D; figure 7) also continuing their spherical growth in volume, as observed by Sciamma'O-Brien et al. (2017).

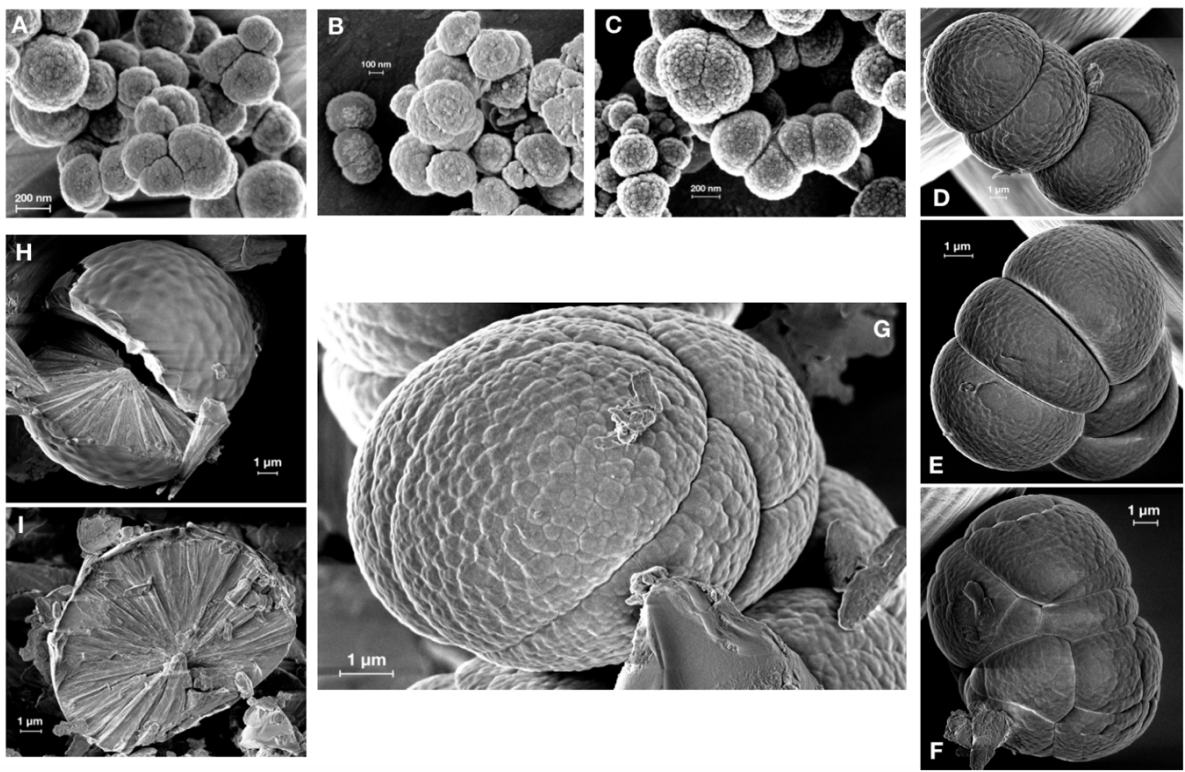

*Figure 7 - SEM images showing specific, repetitive monomer morphologies. Images A, B and C: Coagulation of monomers with plasma residence times limited to 50s, 100s and 145s respectively. Images D, E, F and G: Monomer aggregates formed with a residence time greater than 145s (continuous production). Images H and I: Internal structures of spherical monomers formed with the longest residence time of our experimental conditions (Continuous production).*

This coagulation process appears to repeat itself regularly, involving spherical monomers with similar diameters and greater volume with increasing growth duration. The coagulation of these monomers leads to the formation of aggregates that appear to systematically revert to a spherical geometry when the monomers have a diameter of less than a micrometer. From micrometer diameter upwards, monomers appear to structure aggregates whose shape may evolve away from a sphere. Coagulation appears to result in some volume growth of solid monomers. We also note that the formation of aggregates is increasingly observed with increasing aerosol residence time within the plasma. When the gas phase becomes stable, the presence of several generations of aerosols in the plasma may further promote coagulation between the solids, leading in part to the micrometric volume growth observed on monomers produced with a continuous plasma discharge.

During observations of the Pul20s production, a sub-micrometric monomer from the aerosol analogues population, which must have resided in the plasma for more than 20s, shows an internal structure in which radially arranged sub-spherical particles of smaller size seem to distinguish (image bottom left of figure 6). Micrometric spherical monomers from the Continuous production also reveal a radial internal structure that appears more compact (images H and I in figure 7) than previously observed (Hadamcik et al., 2009; Pul20s production in this study). At this stage of growth, the internal arrangement of individual monomers makes it impossible to distinguish the monomers considered less evolved (smaller diameters) that have coagulated.

## III.c - Analysis of the chemical composition of Titan aerosol analogues: temporal evolution

Figure 8 shows the mass spectra obtained for Titan aerosol analogues produced with plasma discharge times of 20s (Pul20s, figure 8.A), 50s (Pul50s, figure 8.B) and continuously (Continuous, figure 8.C). These spectra, in the restricted m/z range from 100 to 600, show an evolution in the distribution of detected *m/z* ratios (molecules) with increasing discharge time. For Pul20s aerosols (figure 8.A), the distribution of detected molecules is predominantly located around an *m/z* of 200. For aerosols residing 30 seconds longer within the plasma (Pul50s; figure 8.B), the majority distribution remains around 200, with the appearance of a second minority distribution standing out at a higher *m/z* (around 400). For aerosols produced with a continuous plasma discharge (with a residence time greater than 145 seconds; figure 8.C), these still show two distributions of molecules, the first remaining predominantly around *m/z* 200 and a second minor one around *m/z* 400. However, the intensities measured on the spectrum in figure 8.C are lower than for aerosols produced with a controlled plasma discharge (figures 8.A and 8.B).

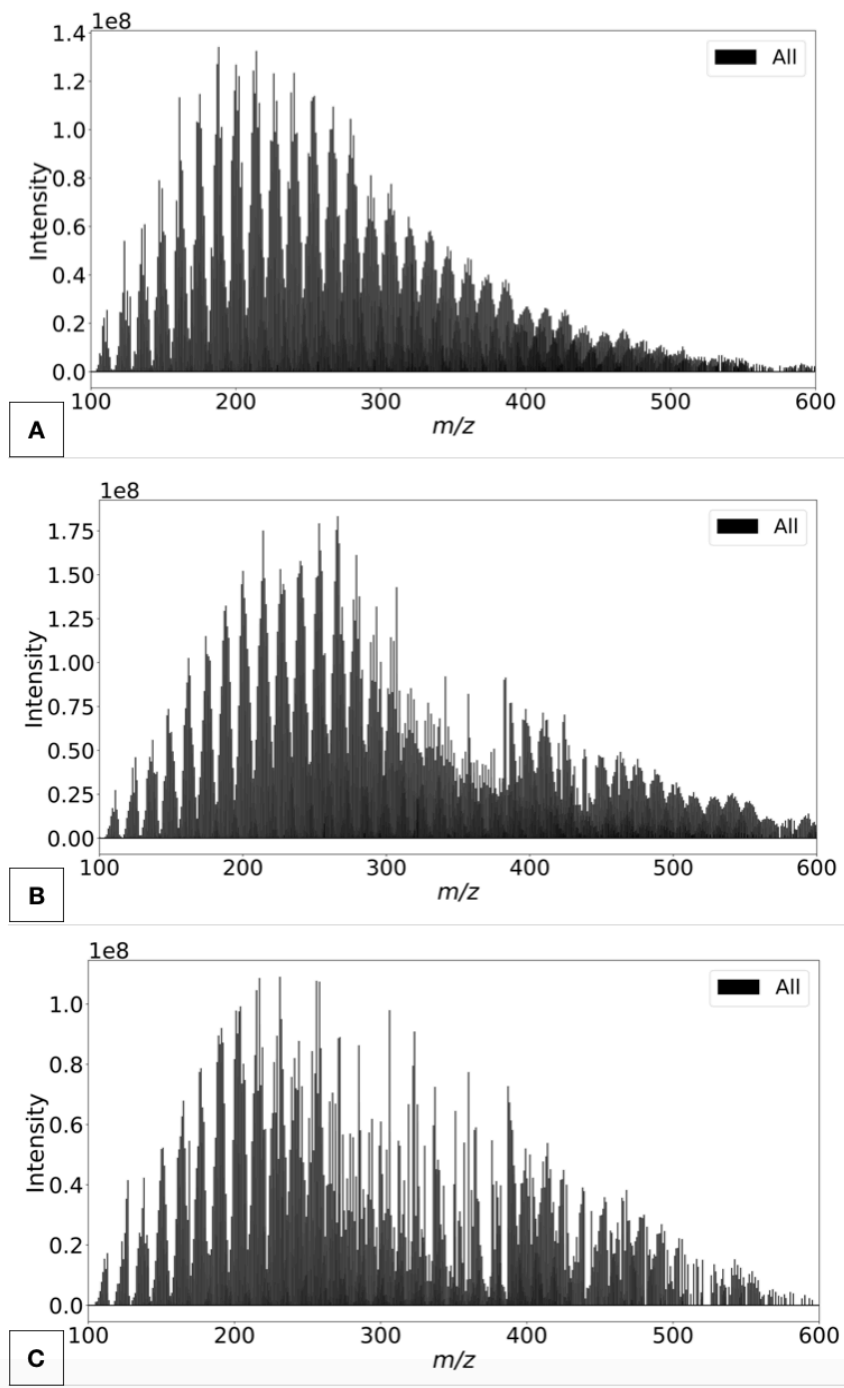

*Figure 8 - Mass spectra obtained for different Titan analog productions. (A): Production Pul20s (discharge duration limited to 20 seconds). (B): Pul50s production (discharge duration limited to 50 seconds). (C): Continue production (continuous discharge, duration of aerosol growth unknown).*

By producing Titan's aerosol analogues using the PAMPRE reactor, chemical analyses of the solids by Gautier et al. (2014) show a variation in the distribution of the molecules detected depending on the experimental conditions set during production, where for the analogues with a nitrogen enrichment, a structure with an intense, isolated distribution positioned mainly around *m/z* 150 is observed. In the *m/z* range between 100 and 600, all the analogues produced during this study show a distribution of molecules predominantly positioned around *m/z* 200, and where the compounds detected predominantly contain at least one nitrogen atom in their assigned crude formulae (pure hydrocarbons rare or non-existent in Continuous production; table 6), correlating with what has been observed previously.

In this same *m/z* range (100 - 600), the number of compounds detected can vary by half between certain aerosol productions (table 6). The first generations of monomers evolved up to a duration of 50 seconds, contain the highest number of compounds (around 4,000). From 50 seconds onwards, the number of compounds detected tends to decrease (table 6), until it is halved (to around 2200) within the monomers formed with the longest estimated residence time, correlating with the drop in intensities measured for each *m/z* detected (figure 8.C). Up to a duration of around 50 seconds, solid monomers possess the greatest diversity of molecules detected, accompanied in parallel by an increase in solid aerosol production (calculated average number of solid monomers present $N_{Part,moy}$; figure 5.C). After 50 seconds and up to an unknown duration, the molecules detected within the solids tend to concentrate around specific *m/z* with the appearance of a second minority distribution of heavier solid constituents, where in parallel the $N_{Part,moy}$ is increasing at a much slower pace. These initial observations of the time-dependent evolution of the chemical composition of Titan's aerosol analogues appear to be temporally consistent with the evolution of the gas phase kinetics observed previously. The gas kinetics of products such as $C_2H_2$, $C_2H_6$ and $HC_3N$ up to 45-50s, show an increase of their abundances with the highest average experimental reaction rate constants $k_{Prod}$ calculated (figure 3, table 3). The prime kinetics of hydrocarbons during the first steps of the transient regime suggests a dominant role of gaseous hydrocarbons in solid core formation mechanisms. The combined action of HC3N confirms a necessary participation of gaseous nitrogen species within solid formation mechanisms. $HC_3N$ appears as a good candidate to associate with light hydrocarbons in copolymerization reactions, as shown by Gautier et al. (2014).

**Table 6 -** Number of compounds detected within the different Titan analog aerosols produced with different residence times within the plasma, as well as the Percentage (%) in these detected compounds containing at least one nitrogen atom in its assigned gross formula. Data are filtered, where any constituent containing oxygen and with an *m/z* greater than 600 are not taken into account.

For information purposes, the number of molecules detected (without oxygen) is given together with the number of additional carbon molecules detected with at least one oxygen atom (*$O_x$) and that of oxygenated molecules comprising nitrogen (**$N_xO_x$) which have been removed in the treatments presented.

| Experimental Conditions | | FTICR-LDI-MS: *m/z* range from 100 to 600 | |
|---|---|---|---|
| Production names | Plasma discharge time (s) | Number of compounds detected in aerosols | % Compounds detected containing at least one nitrogen atom ($C_xH_yN_z$) |
| Pul20s | 20 | 4014 ± 400* / 3573** | 99.63 |
| Pul35s | 35 | 4151 ± 346* / 2769** | 99.73 |
| Pul50s | 50 | 4099 ± 320* / 3434** | 99.89 |
| Pul75s | 75 | 3727 ± 307* / 3003** | 99.89 |
| Pul100s | 100 | 3309 ± 262* / 3207** | 99.96 |
| Pul145s | 145 | 3442 ± 296* / 3791** | 99.96 |
| Continuous | continue | 2247 ± 209* / 3392** | 100.0 |

To better understand the evolution of the chemical composition of the aerosol analogues, figure 9 shows the distribution of the different classes of $C_xH_yN_z$ heteroatoms detected in each production of solids, taking into account the relative intensities measured at each *m/z* detected and associated with a $C_xH_yN_z$ compound. For aerosols produced with a discharge duration of less than 50 seconds (Pul20s, Pul35s), the most abundant nitrogen compounds detected contain 4 (N4) or 3 (N3) nitrogen atoms, with relative intensities of around 17% for Pul20s production. Compounds containing more than 6 nitrogens are in the minority, with abundances below about 3 - 4%, and even below 1% for compounds containing 10 or more nitrogens in Pul20s aerosol analogues. We note that when the aerosols have resided for a further 15 seconds (Pul35s), the relative intensities of majority compounds with N3 or N4 begin to decrease (becoming below 16%), in contrast to compounds containing N5 or N6 nitrogens, where the relative intensity characterizing the presence of the N6 heteroatom class increases in parallel. When the duration of the discharge exceeds 50 seconds (from Pul50s production), the majority of compounds become N6 and N5. As plasma discharge duration increases from Pul50s to Pul145s, compounds containing more than 6 nitrogens (N7 to N15) become increasingly abundant, in contrast to compounds containing 4 or fewer nitrogens (N4 to N1), see for aerosols produced with the longest estimated growth time (Continuous production), compounds containing 9 (N9) and 10 (N10) nitrogens become among the most abundant. We therefore observe that the N-content of the aerosols analogs increases after about 50s. It is interesting to note thate there might be a correlation between the second phase of $CH_4$ consumption and the delayed nitrogen molecule formation observed in the gas phase.

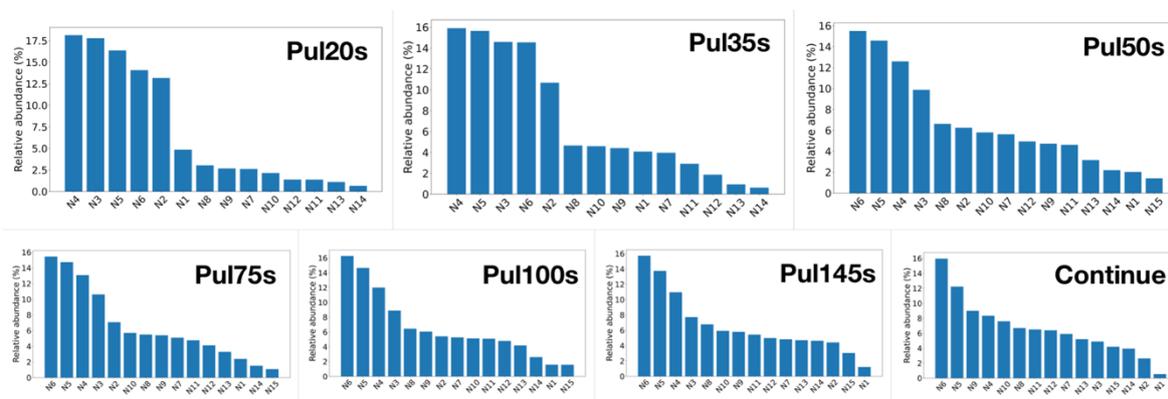

*Figure 9 - Distributions of $C_xH_yN_z$ heteroatom classes detected within each Titan analog production (table 1). The molecular formulas taken into account have an intensity value greater than 0.5% of the most intense measured.*

Gautier et al. (2014) also show a variation in nitrogen incorporation in their plasma-produced Titan aerosol analogues, injecting initial $CH_4$ proportions of 1 and 10% with a higher gas flow (55 sccm). Gautier et al. (2014) detect similar compounds between analogues produced with different initial $CH_4$ proportions. However the predominant compounds differ, notably depending on whether they contain 6 (N6) or 4 nitrogen atoms (N4). With 1% $CH_4$ initially injected, the more nitrogen-enriched N6 compounds are more abundant, correlating with stable gas chemical composition analysis which shows a predominance of nitrogen products over hydrocarbons (Sciamma'O-Brien et al., 2010; Carrasco et al., 2012; Gautier et al., 2014). Whereas by injecting 10% initial $CH_4$, N4 compounds predominate the solid this time, correlating with a predominance of gaseous hydrocarbons over nitriles, optimizing carbon rather than nitrogen enrichment in analogous aerosols (Sciamma'O-Brien et al., 2010; Carrasco et al., 2012; Gautier et al., 2014; Derenne et al., 2012). Here, the first generations of monomers (Pul20s) and the least evolved (Pul35s) exhibit dominant N4 and N3 compounds, in tune with a more active participation of hydrocarbons during solid copolymer formation processes. Pul50s, Pul75s, Pul100s, Pul145s and Continuous show nitrogen enrichment with predominant N6 compounds, where in parallel from 50 seconds onwards, gaseous hydrocarbons enter a kinetic sub-phase characterizing a consumption with a rate constant $k_{Cons}$ still higher than that of gaseous nitriles (table 2), who themselves are still in their kinetic sub-phases of production until around 70 seconds. From this point onwards, hydrocarbons are likely to become much less abundant than nitriles, giving way to gaseous nitrogen molecules as the dominant player in the aerosol analogues growth processes that take place after 50 seconds.

Increasing the growth time of solid monomers has enabled the chemical evolution of compounds with some incorporation of nitrogen. Compounds containing N3, N4, N5 and N6 nitrogen atoms are present from the very first generation of monomers (Pul20s), as well as within the most evolved monomers, however the variation in their abundances over time may reflect different chemical or physical processes of solid formation and growth. Compounds containing more than 6 nitrogen atoms appear also to be formed by other chemical growth processes/other chemical growth step. To better understand the chemical evolution of the aerosol analogues produced in this study, Van Krevelen diagrams are used to represent the molecular complexity of Titan aerosol analogues (Pernot et al., 2010; Imanaka and Smith, 2010; Gautier et al., 2014; Maillard et al., 2018). Based on the molecular formulas, the N/C ratios are plotted as a function of the H/C ratio in figure 10.

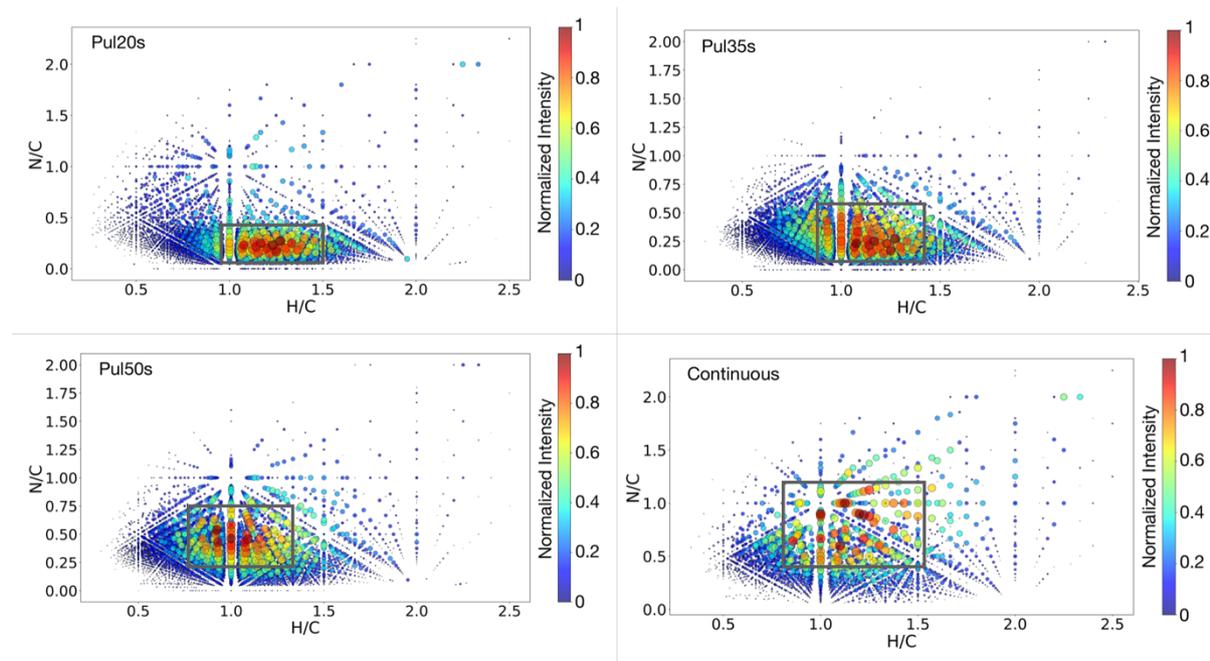

*Figure 10 - Inverted Van Krevelen diagrams showing N/C ratios as a function of H/C ratio, determined from the molecular formulas assigned to the different Titan analog productions. The majority molecular distribution detected is framed and represented by the red color code. Pul20s production means a production with a discharge duration limited to 20 seconds. Pul35s production with discharge duration limited to 35 seconds. Pul50s production with discharge time limited to 50 seconds. Continuous production with continuous discharge (duration of aerosol growth unknown).*

For Pul20s and Pul35s productions (figure 10), the majority of molecules are distributed with an N/C ratio below 0.5, and an H/C ratio between 1 and 1.5, indicating that the molecules making up the solids are rather unsaturated (linear or branched) and enriched in carbon and hydrogen. However, by increasing the discharge time between these two productions (Pul20s and Pul35s), the distribution of molecules evolves with an N/C ratio increasing between values of 0.25 and 0.5, and an H/C ratio migrating predominantly around 1. For a duration of 50 seconds (figure 10), the molecules show an enrichment in nitrogen with an N/C ratio settling around 0.5, as well as a decrease in hydrogen marked by an H/C ratio settling around 1. This transition can be expressed by a greater amount of unsaturation in the solid constituents. For aerosol analogues productions carried out with a discharge duration of 75 to 145 seconds (Pul75s, Pul100s, Pul145s), the compounds show a similar trend to those of Pul50s (figure 10), with an H/C ratio set at 1 and an N/C ratio around 0.5 migrating slightly towards 0.7. Aerosol analogues with the longest growth times (Continuous, figure 10) continue to enrich in nitrogen, with species distributed around an N/C ratio of 1 and weighted by the H/C ratio remaining close to 1 (with no further loss of hydrogen). In the case of the aerosols analogues produced in continuous mode, the types of molecules seem to be more spread out, suggesting a broader diversity of molecules. This difference confirms that the first steps of the aerosols formation occurring during the transient regime are more specific than the evolution process occurring during the stationary state.

In the case where 10% initial $CH_4$ is injected into the PAMPRE reactor, the solid compounds produced in the plasma have an N/C ratio of less than 0.5 (Gautier et al., 2014) and an increase in unsaturated hydrocarbon content observed by nuclear magnetic resonance (Derenne et al., 2012), where in parallel the presence of hydrocarbons predominates the stabilized/final gas phase (Gautier et al., 2011). With 5% initial $CH_4$, nitrogenous gaseous products are dominant when the composition becomes stable (similar case of 4% from Gautier et al., 2011), and aerosol analogues produced in plasma have constituents predominantly distributed around m/z ratios between 400 - 450, comprising 5 (N5) to 6 (N6) nitrogen atoms, with an N/C ratio around 0.5, or even extending to 1 by decreasing the initial proportion of $CH_4$ to 1% (Gautier et al., 2014). By injecting at 55 sccm initial $CH_4$ proportions ranging from 1 to 10%, Wattieaux et al. (2015) measured the self-biasing voltage (Vdc) of the powered electrode and observed that the initial proportion and consumption efficiency of $CH_4$ influence the ionic chemistry taking place within the plasma. They observed that an increase in the initial proportion of $CH_4$ leads to an increase in the initial proportion of ions, facilitating the ionization of $CH_4$ and $N_2$. Yet an efficient $CH_4$ consumption optimizes ion production, particularly of light positive species such as $N_4^+$ as shown by Alves et al. (2012). Between 1 and 5% initial $CH_4$, $CH_4$ consumption efficiency is considered optimal at a gas flow rate of 55 sccm (Sciamma'O-Brien et al., 2010), and ion production appears to be as well (Wattieaux et al., 2015), so nitrogen enrichment in solids can be promoted by ion chemistry (Gautier et al., 2014). On the opposite, it was shown that without the support of ion chemistry, aerosol analogues produced from a copolymerization process of neutral gases extracted from the PAMPRE experiment are lighter and predominantly comprise 2 (N2) to 4 (N4) nitrogen atoms. (Gautier et al. (2014)). In the latter case, the gas phase products had been produced with standard plasma conditions in the PAMPRE reactor using a 5% methane reactive gas mixture and cumulated in a cryogenic trap.

Despite an increase in the initial proportion of $CH_4$ to 20% and the proportions of $CH_4$ consumed, we observe correlations between our aerosol analogues at different stages of growth, and the observations made by previous studies injecting 1 to 10% initial $CH_4$ (summarized in table 7). The chemical composition of the first generation of aerosols (Pul20s) seems to testify to a dominant participation of neutral gaseous hydrocarbons in the chemical pathways of solid formation, as for the case with 10% initial $CH_4$ (table 7). The chemical composition of more advanced analogues (Pul35s) shows a nitrogen enrichment during solid growth, certainly involving gaseous nitrogenous players, as for the cases with 1 and 5% initial $CH_4$ (table 7). Here, with a higher initial $CH_4$ proportion and and its efficient consumption, the gaseous ionisation in the plasma may be favoured. Ionic nitrogen products are therefore involved in the incorporation of nitrogen into the solid up to Pul50s (with a loss of hydrogen). After 50s, the incorporation of additional nitrogen may involve other potentially neutral, radical or ionic gaseous. Indeed, the N/C ratio has been shown to be highly sensitive to experimental conditions, including the chemical processes dominating the molecular growth of aerosol analogues. Imanaka and Smith (2010) have produced Titan aerosol analogues after the irradiation of $N_2$-$CH_4$ gas mixtures by extreme ultraviolet photons. These aerosols exhibit N/C and H/C ratios equally weighted at 1. The authors therefore suspect that a neutral product like HCN may be an important gaseous precursor, intervening in co-polymeric processes with other groups such as -CN or -H or -$NH_x$ or -$CH_x$, in agreement with Gautier et al. (2014) for Titan analogues produced by the PAMPRE experiment. In our experiments, HCN could also be a good candidate to participate in the most advanced solid monomer growth processes (Continuous production), as we observe that a rather large amount of HCN is consumed before gas concentrations stabilize in the plasma (table 3).

**Table 7 -** Comparison of the chemical compositions of the gas phase and aerosol analogues formed using different experimental conditions set in the PAMPRE dusty plasma reactor.
*data from Sciamma'O-Brien et al. (2010) / **data from Wattieaux et al. (2015) / *** data from Gautier et al. (2014)

| Chemical composition of gas phase | | | | Chemical composition of aerosol analogues | |
|---|---|---|---|---|---|
| % $CH_4$ initial and Gas flow injection | | % $CH_4$ consumed | | Total duration of transient $CH_4$ consumption kinetics (s) | Ratio of solid constituents detected*** | |
| | | | | | N/C | H/C |
| 1* | at 55 sccm | 0.82 | at stable gas phase | 40 | 1 | 1.5 |
| 5** | | 3.35 | | 75 | 0.5 | 1.5 |
| 10* | | 4.47 | | 90 | < 0.5 | 1.5 |
| 20 (This work) | at 2.5 sccm | 5.54 | at 20s | stop at 20 | < 0.5 | > 1 |
| | | 12.84 | at 50s | stop at 50 | 0.5 | 1 |
| | | 19.41 | stationnary state | 150 | 1 | 1 |

The loss of hydrogen between 20 and 50 seconds, with the H/C ratio migrating from 1.5 to settle at around 1, reveals a process taking place during the incorporation of nitrogen into the solid constituents. A change in the molecular structure of the solid constituents (aromaticity of the molecules) through catalytic reactions at aerosol surface could explain this loss of hydrogen (Yung et al., 1984; Courtin et al., 1991; Bakes et al., 2003; Imanaka et al., 2004). To better understand the arrangement of atoms within solid constituents during the observed incorporation of nitrogen, the number of unsaturations within the detected compounds is represented by the calculation of the DBE (Double bond equivalent according to $DBE = C - \frac{H}{2} + \frac{N}{2} + 1$) as a function of the number of carbon C, on the maps in figure 11. On these maps, the polycyclic aromatic hydrocarbons (PAH) line is added in red, representing the least saturated family of species containing only carbon and hydrogen (Cho et al., 2011).

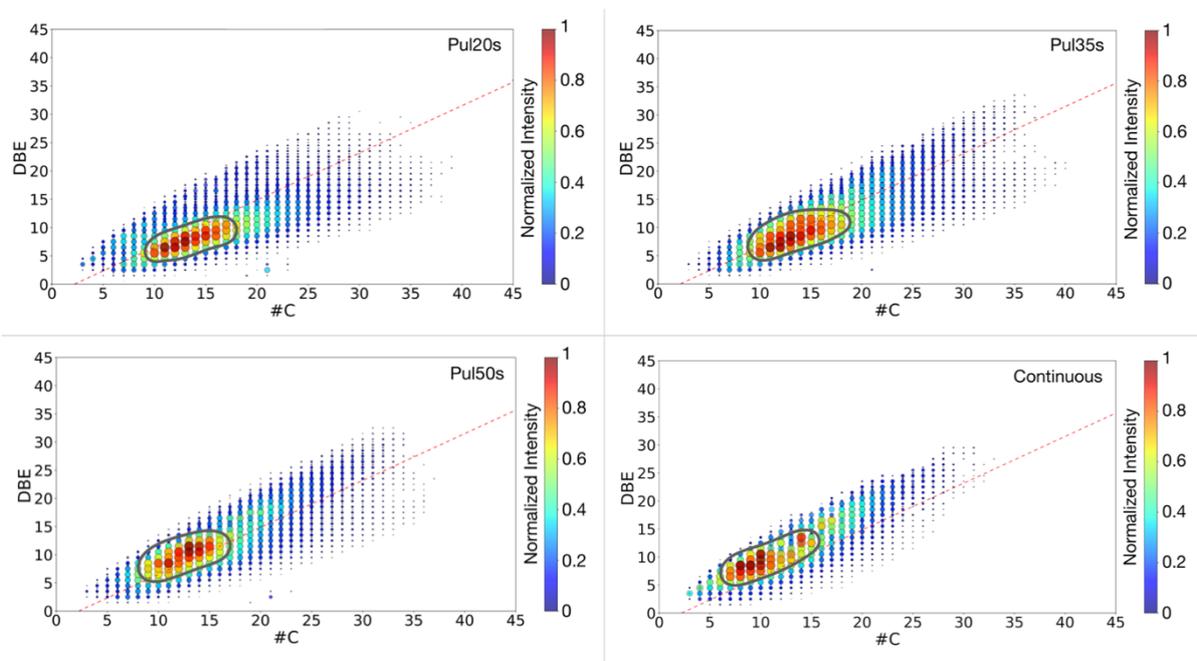

*Figure 11 - Double bond equivalent (DBE) mappings as a function of the number of C atoms determined from the molecular formulae assigned to the different Titan analog productions. The majority molecular distribution detected is circled and color-coded red. The red line represent the PAH line. Pul20s production with discharge duration limited to 20 seconds. Pul35s production with discharge duration limited to 35 seconds. Pul50s production with discharge time limited to 50 seconds. Continuous production with continuous discharge (duration of aerosol growth unknown).*

For the first generation of monomers (Pul20s, figure 11), the majority distribution of molecules has an unsaturation number versus C number below the PAH line, revealing the presence of predominantly aliphatic molecules (branching processes within the solid constituent). As shown by Carrasco et al. (2012), atomic hydrogen is abundant within the gas phase simulating Titan's atmospheric chemistry, using the PAMPRE dusty plasma reactor (fragmentation of hydrocarbons). They demonstrated an inhibitory effect of atomic hydrogen due to its efficient attachment to the surface of organic solids, to the detriment of solid growth and production. Sekine et al. (2008.a; 2008.b) have shown that the release of atomic hydrogen during the fragmentation of $CH_4$ gas can lead to collisions with the surface of forming and growing solid aerosols within the plasma, leading to faster hydrogenation than the hydrogen abstraction achieved by heterogeneous reactions. During the Pul20s production of this study, hydrocarbons and $CH_4$ exhibit gaseous kinetics with a significant reaction rate constant. Their fragmentation can produce an important concentration of atomic hydrogen in the discharge in the first steps of the transient regime. We therefore suspect that a hydrogenation process also contributes to limit the number of unsaturations within the molecules composing the solid in the Pul20s, favoring linear or branched arrangements with a significant hydrogen enrichment (1< H/C < 1.5) compared to the other solid productions produced with longer pulses. The number of solid monomers produced does not appear to be affected, but the growth of the particles seem limited/constrained, compared with that observed on aerosol analogues residing longer in the plasma (figures 5.C and 5.A respectively).

With increasing discharge time, the majority distribution of molecules has an increasing number of unsaturations (Pul35s, figure 11), until it exceeds the PAH line (Pul50s, figure 11). During this period, the increase in DBEs may be due to the incorporation of nitrogen previously observed in the molecules, as well as the loss of hydrogen which would also begin within the monomers of the Pul35s production. This loss of hydrogen would initiate a phenomenon of cyclization of the solid constituents until they become aromatic. This chemical evolution takes place after a significant enrichment of HCN in the gas phase (figure 2), the presence of which could promote the incorporation of nitrogen into the organic monomers, forming nitrogen-containing polycyclic aromatic compounds according to Imanaka et al. (2004) and Perrin et al. 2021. In addition, collisions between electrons and ions/radicals on the surface of Titan analogs within the cold plasma can break aliphatic C-H bonds, removing hydrogen from the solid surface and forming stable aromatic rings within the solids, as shown by Imanaka et al. (2004). For Continuous production (figure 11), the DBE value stabilizes at around 8-9, with an average of 10 C atoms present or less (distribution above the PAH line). These aerosols with the longest growth times show additional nitrogen enrichment taking place within the aromatic compounds. As shown by Imanaka et al. (2004), HCN can also enter aromatic carbon ring networks located within cold plasma-produced Titan analogues.

### III.d - Determination of uptake coefficients:

The kinetics of HCN gas observed in our experiments lead us to suspect that the surface of growing aerosols could serve as gas adsorption sites, but also for other neutral products ($C_2H_2$, $C_2H_6$, $HC_3N$, $C_2H_3N$ or $C_2N_2$). Organic solids would act as sinks for neutral gas consumption within the plasma, through heterogeneous processes such as adsorption, but also chemical reactions, desorption and absorption. These heterogeneous surface processes could also affect the microphysical growth mechanisms observed on our analogues, by realizing a surface chemistry.

The experimental rate reaction constants k determined for various neutral gases are not directly extrapolable to natural study cases such as Titan's atmosphere. For this reason, so-called uptake coefficients **γ** are more appropriate to describe the kinetics of heterogeneous processes between neutral gas products and organic aerosols (Crowley et al., 2010 ; Pöschl et al., 2007). As described by IUPAC (Subcommittee for Gas Kinetic Data Evaluation), relationship (7) is widely used to describe coefficient **γ** as the net probability that a gaseous molecule X undergoing a kinetic collision with a solid surface will actually be absorbed at the surface. This approach links processes at the interface and beyond to an apparent first-order loss of X from the gas phase.

$$\frac{d[X]_g}{dt} = -k_{exp}[X]_g = -\gamma \frac{\bar{c}}{4}[SS]_g[X]_g \quad (7)$$

where $[SS]_g$ is the specific surface area represented by the solid phase as a function of the volume of the gas phase ($m^{-1}$). In our case, $[SS]_g$ corresponds to the measured geometric surface area of a spherical aerosol (with $d_{moy}$ from table 3, $m^2$) multiplied by the calculated average number of aerosols present ($N_{Part,moy}$; figure 5.C), all related to the total volume of the cage confining the plasma (plasma cage with a cylindrical volume of $3.92.10^{-4}$ $m^3$). $\bar{c}$ is the average thermal velocity of gaseous species X ($m.s^{-1}$). $k_{exp}$ is a constant characterizing a first-order loss of species X in the gas phase ($s^{-1}$). In our case, $k_{exp}$ corresponds to the average $k_{Cons}$ calculated to characterize the kinetic sub-phase of gaseous product consumption (table 3).

In our study, we carried out 3 productions of aerosol analogues that had completed their growth in parallel with the kinetic sub-phase of gas consumption (Production Pul75s, Pul100s and Pul145s). The necessary characteristics ($N_{Part,moy}$ calculated with $\rho_{aerosol}$ = 0.4 and 1.44 $g.cm^{-3}$ in (5), $d_{moy}$ of table 4) determined for these 3 productions were used to calculate the specific surface area $[SS]_g$ represented by the solids at different durations of gas consumption (75, 100, 145 s), and are summarized in table 8. It can be seen that as the duration of the plasma discharge increases, the specific surface area represented by the solid as a function of the volume of the gas phase (gas-solid interaction volume) also increases, but remains of the same order of magnitude ($10^{-2}$ cm$^{-1}$).

In our study, we carried out 3 productions of aerosol analogues that had completed their growth in parallel with the kinetic sub-phase of gas consumption (productions Pul75s, Pul100s and Pul145s). With the two densities ($\rho_{aerosol}$ = 0.4 and 1.44 g.cm$^{-3}$ in relation (5)) from the literature fixed to the solids for the same analogues production, the average number of aerosols present within the plasma over the fixed plasma discharge duration (N$_{Part,moy}$) varies by an order of magnitude (table 8). By fixing the average diameter d$_{moy}$ of a monomer/aerosol determined for each production run (section III.b.a, tables 4 and 8), the two N$_{Part,moy}$ values were used to calculate the specific surface area [SS]$_g$ represented by the solids at different gas consumption times (75, 100, 145 seconds), and are summarized in the table 8. For the two N$_{Part,moy}$ values used, it can be seen that as the duration of the plasma discharge increases, the specific surface area represented by the solid as a function of the volume of the gas phase (gas-solid interaction volume) increases. For Pul75s production, the value of [SS]$_g$ remains close despite the variation in N$_{Part,moy}$, but for Pul100s and Pul145s production, [SS]$_g$ varies by an order of magnitude, going from $10^{-1}$ cm$^{-1}$ to $10^{-2}$ cm$^{-1}$ when N$_{Part,moy}$ decreases from $10^{+10}$ to $10^{+9}$ respectively.

Taking into account the two values of [SS]$_g$ where a different density is assigned to the aerosols of each analogues production, for the 6 neutral gaseous species analyzed previously, table 8 summarizes the calculated experimental uptake coefficients γ. Note that when only the geometric surface area of the particles is taken into account to determine the specific surface area [SS]$_g$, the calculated values of γ correspond to an upper limit (IUPAC, 2007). With this study, the first experimental determinations of uptake coefficients of neutral gaseous molecules were carried out for $C_2H_2$, HCN, $C_2H_6$, $C_2H_3N$, $HC_3N$ and $C_2N_2$ on Titan aerosol analogues, mimicking gas-particle interactions in a planetary atmosphere. The uptake coefficients determined are all of the order of $10^{-4}$ - $10^{-5}$, and the presence or absence of nitrogen in the gaseous molecule does not appear to be a decisive factor in bonding efficiency. We note that for Pul75s production, γ values remain in the order of magnitude of $10^{-4}$ for each gaseous species, with the exception of HCN and $C_2H_3N$ whose order drops to $10^{-5}$ with decreasing [SS]$_g$ (table 8). For Pul145s production, a similar behavior is observed, where for the γ value calculated for all gases presents an order of magnitude of $10^{-5}$ (table 8). For Pul100s, γ shows a variation of one order of magnitude, going from $10^{-4}$ to $10^{-5}$ as the value of [SS]$_g$ decreases from $10^{-2}$ cm$^{-1}$ to $10^{-1}$ cm$^{-1}$ respectively (table 8). It is also observed that for each condition, whether by varying the [SS]$_g$ value for the same production, or between the different productions, the γ value determined for $HC_3N$ is always the highest between the 6 gaseous products analyzed (table 8). Normalizing the γ values of each gas by the correlating γ$_{HC3N}$, we observe that the γ are similar between the 6 cases for each individual product: $C_2H_2$ (0.89), HCN (0.45), $C_2H_3N$ (0.65), $C_2N_2$ (0.78) and $HC_3N$ (1). Similarly, the higher average γ value calculated by taking into account the calculated values of the three aerosol analogues productions is of the same order of magnitude between each gaseous product analyzed for a value set at [SS]$_g$ (table 8). Based on these observations, uptake coefficients appear to be generally independent of the concentration and size of aerosol analogues, or the velocity of the surrounding gas. They reflect properties intrinsic to gas-particle interaction and can therefore be used to extrapolate the process to other conditions, such as Titan's atmosphere.

Table 8 - Experimental uptake coefficients γ determined between aerosol analogues and different neutral gaseous species, carried out using the PAMPRE experiment. γ is calculated from relation (7), and the experimentally determined physical characteristics ($k_{Cons}$ from table 2; mean diameters from table 3; $N_{Part,moy}$ from figure. 5.C). The uncertainties on $[SS]_g$ are given by a quadratic sum, taking into account the average uncertainty determined on the value of the average diameter retained for the monomers and the uncertainty determined on the average calculated number of monomers present $N_{Part,moy}$. The uncertainties on γ are given by the sum of the uncertainties obtained for $[SS]_g$ and $k_{Cons}$.
with $\rho_{aerosol}$ = 0.4 g.cm$^{-3}$ / 1.44 g.cm$^{-3}$ in relation (5).

| Aerosol production | Average aerosol diameters (nm) | Average number of aerosols present $N_{Part,moy}$ | | Specific surface area $[SS]_g$ (cm$^{-1}$) | |
|---|---|---|---|---|---|
| Pul75s | 483 ± 16 | (2.4 ± 0.2).10$^{+9}$ | (6.7 ± 2.4).10$^{+8}$ | (4.53 ± 0.04).10$^{-2}$ | (1.26 ± 0.40).10$^{-2}$ |
| Pul100s | 408 ± 12 | (1.0 ± 0.1).10$^{+10}$ | (3.1 ± 1.0).10$^{+9}$ | (1.36 ± 0.13).10$^{-1}$ | (3.77 ± 1.33).10$^{-2}$ |
| Pul145s | 509 ± 18 | (1.1 ± 0.1).10$^{+10}$ | (2.9 ± 1.1).10$^{+9}$ | (2.34 ± 0.23).10$^{-1}$ | (6.50 ± 2.29).10$^{-2}$ |

| Aerosol production | Uptake coefficient γ ( ± ).10$^{-5}$ / ( ± ).10$^{-4}$ | | | | | |
|---|---|---|---|---|---|---|
| | $C_2H_2$ | | HCN | | $C_2H_6$ | |
| Pul75s | 12.6 ± 2.5 | 4.6 ± 1.8 | 6.5 ± 1.9 | 2.3 ± 1.0 | 12.3 ± 2.8 | 4.4 ± 1.8 |
| Pul100s | 4.2 ± 0.8 | 1.5 ± 0.6 | 2.2 ± 0.6 | 0.8 ± 0.3 | 4.1 ± 0.9 | 1.5 ± 0.6 |
| Pul145s | 2.5 ± 0.5 | 0.9 ± 0.3 | 1.2 ± 0.4 | 0.4 ± 0.2 | 2.4 ± 0.5 | 0.9 ± 0.4 |
| Higher average γ | 6.4 ± 1.3 | 2.3 ± 0.9 | 3.3 ± 0.9 | 1.2 ± 0.5 | 6.3 ± 1.4 | 2.3 ± 0.9 |
| Normalized γ | 0.89 | | 0.45 | | 0.87 | |
| | $C_2H_3N$ | | $HC_3N$ | | $C_2N_2$ | |
| Pul75s | 9.2 ± 2.1 | 3.3 ± 1.4 | 14.2 ± 2.9 | 5.1 ± 2.0 | 11.1 ± 2.5 | 4.0 ± 2.3 |
| Pul100s | 3.1 ± 0.7 | 1.1 ± 0.5 | 4.7 ± 0.9 | 1.7 ± 0.7 | 3.7 ± 1.7 | 1.3 ± 0.8 |
| Pul145s | 1.8 ± 0.4 | 0.6 ± 0.3 | 2.7 ± 0.5 | 1.0 ± 0.4 | 2.2 ± 0.9 | 0.8 ± 0.4 |
| Higher average γ | 4.7 ± 1.1 | 1.7 ± 0.7 | 7.2 ± 1.5 | 2.6 ± 1.0 | 5.7 ± 2.6 | 2.0 ± 1.2 |
| Normalized γ | 0.65 | | 1 | | 0.78 | |

Between altitudes ranging from the mesosphere to the troposphere (700 - 100 km), CIRS/Cassini observations probing different polar latitudes (northern and southern hemispheres) have measured the mixing ratios of numerous gaseous products, notably $C_2H_2$, HCN, $HC_3N$ and $C_2H_6$. According to Mathé et al. (2020), these mixing ratios vary with altitude, and the profiles calculated for $C_2H_2$ and HCN in particular show variations with a similar pattern in the polar zone located in the northern hemisphere, where an even more abundant presence of aerosols is expected due to convective atmospheric circulation phenomena (Rannou et al., 2010). In our experiments, the shape of the slope and the values of the experimental kinetic coefficients characterizing the loss of gaseous products are also similar for $C_2H_2$ and HCN, for example. Taking into account a loss due to consumption by interactions with solid aerosol analogues, the uptake coefficients determined for these two neutral gaseous products are similar and may play a part in their loss/consumption within the experiments, and potentially also between aerosols and gases atmospheric on Titan.

These experimental uptake coefficients γ are obtained from observed gas species loss rates and calculated collision rates with the geometric surface of solid particles only, without taking into account the involvement of other heterogeneous mechanisms taking place in parallel or the chemical nature of the solids. Indeed, a correct description of the heterogeneous interactions of a trace gas with a solid surface must generally take into account several competing transport (gas-phase diffusion, adsorption, desorption) and surface and bulk chemical reactions (Cronwel et al., 2010; Pöschl et al., 2007; Ammann and Pöschl, 2007). As these processes can work against the absorption of gas by the solid, they can influence the calculated γ values by applying corrections characterizing these heterogeneous interactions and give a lower limit to the γ values.

## IV - Discussions:

### IV.a – Aerosol growth by experimental observations:

Using the PAMPRE dusty plasma reactor, we have temporally sequenced and identified different stages in the formation and growth of aerosol analogues to those present in Titan's atmosphere. Morphological (SEM) and chemical (FTICR-LDI-MS) analyses of the analogous aerosols enabled us to observe a temporal evolution of these properties within our experimental simulation. The evolution observed in organic aerosols shows correlations with the kinetics of certain gaseous products present in the gas phase, such as $C_2H_2$, HCN, $C_2H_6$, $C_2H_3N$, $HC_3N$ and $C_2N_2$, distinguishing their participation in different mechanisms of solid production/formation and growth. The diagram presented in figure 12 combines the various experimental observations made during our experiments simulating the atmospheric organic chemistry of the Titan satellite.

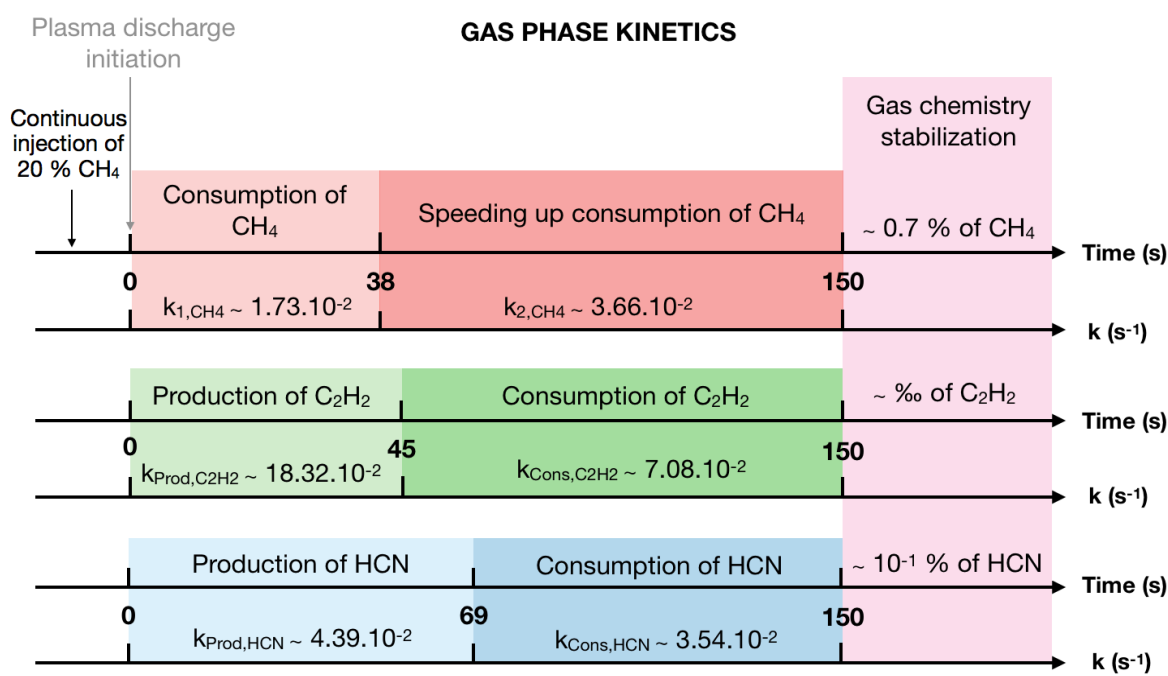

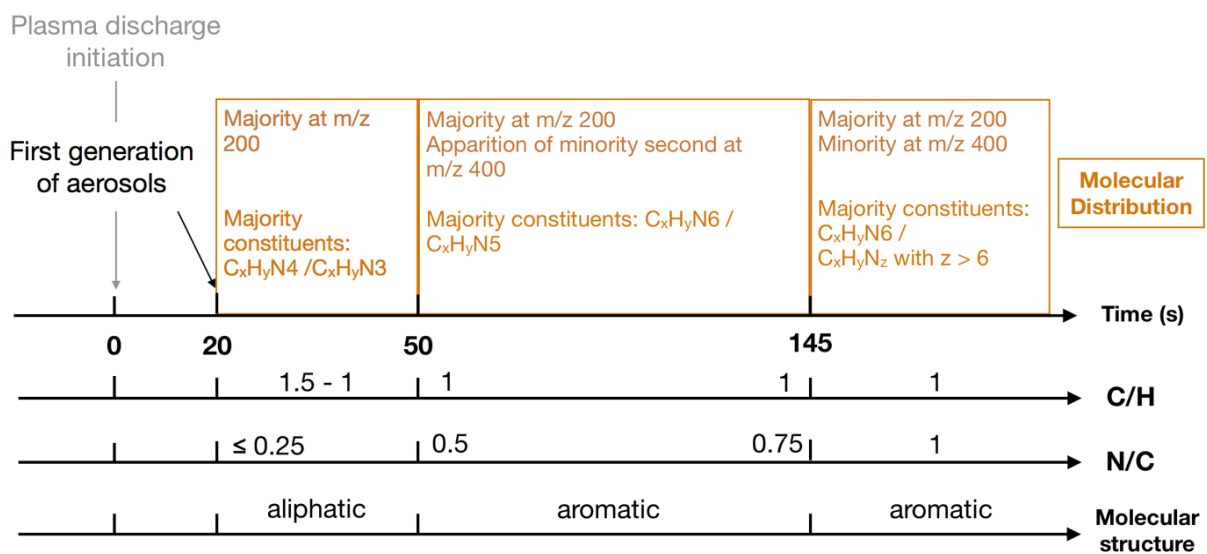

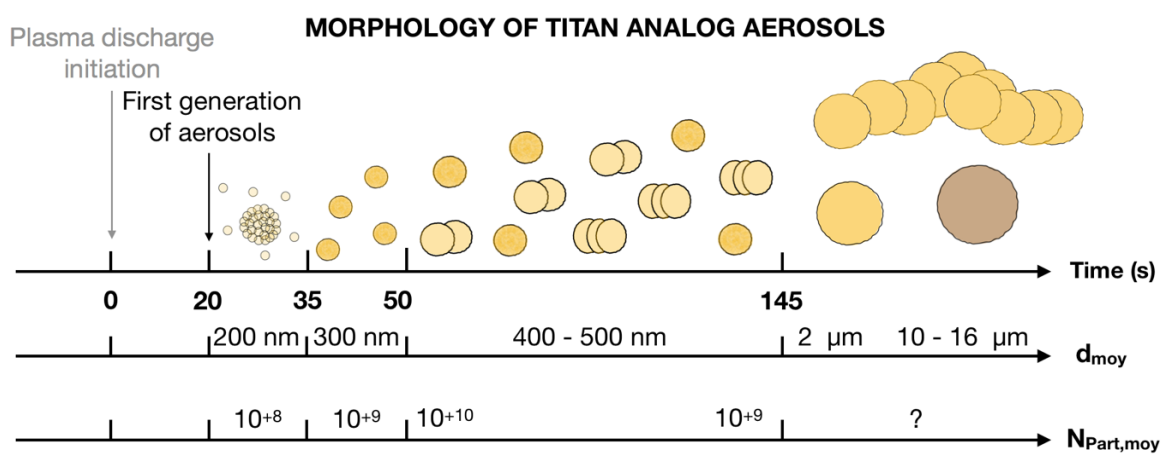

*Figure 12 - Schematic diagram summarizing the microphysical and chemical time evolutions observed for analog aerosols from Titan, and the kinetic evolution of neutral gaseous products occurring in parallel in the PAMPRE dusty plasma reactor.*

Twenty seconds after initiation of the plasma discharge, the first generation of solid aerosols appears, possessing the smallest range of diameters measured (with a $d_{moy}$ ~ 200 nm). As soon as they are formed, these have a chemical composition, where the compounds detected predominantly possess at least one nitrogen atom and are represented by different classes of $C_xH_yN_z$ heteroatoms, the most dominant of which are composed of 3 or 4 nitrogen atoms (called N3 or N4 respectively). The chemical composition of the first generation is also the most enriched in hydrogen (H/C ~1.5) compared to aerosols with a longer growth time (discharge time > 20s), with the particularity of being made up of compounds with an aliphatic structure (ramified).

By increasing the discharge duration to 50 seconds, the production of solids increases with the calculated average number of aerosols present in the plasma ($N_{Part,moy}$) increasing by almost two orders of magnitude. At the same time, monomers produced with a exposure time of more than 20 s, show growth inducing modifications to the morphology and chemical composition initially observed on the first generation of monomers. The general spherical morphology observed on monomers evolved up to 50 seconds increases in volume with a doubling of average diameter ($d_{moy}$ ~ 400 nm). As for their chemical compositions, an incorporation of nitrogen is observed with compounds containing 5 or 6 nitrogen atoms (N5 and N6) becoming dominant, while a loss of hydrogen (migration of the H/C ratio towards 1) takes place in parallel, inducing an increase in the number of unsaturations of solid compounds.

As observed by Alcouffe et al. (2008) and Wattieaux et al. (2015), the appearance of solid aerosols occurs before the composition of the gas phase stabilizes within the plasma. According to our observations, during the transient regime, the production of hydrocarbon molecules is particularly efficient up to 45 - 50 seconds, and their abundance seems to favor the solid production process. Moreover, over the same time period, the experimental reaction rate constants $k_{Prod}$ calculated are much higher for gaseous hydrocarbons than for nitrogenous species, probably testifying to the dominant involvement of hydrocarbons in the initial solid formation mechanisms. This predominance is consistent with a previous study of Berry et al. (2019a), photolyzing a $N_2$ - $CH_4$ mixture with EUV photons and starting with an initial 2% $CH_4$ diluted in $N_2$ gas. They indeed showed by monitoring the gas phase composition that hydrocarbons were produced before unsaturated organic nitrogen compounds.

The chemical pathways leading to the formation of Titan aerosol analogues produced using the experimental condition set in this study, have been attributed to a co-polymerization process involving neutral gaseous products producing compounds with an aliphatic character. The chemical arrangement of the solid compounds reveals the possibility of the participation of unsaturated hydrocarbons such as $C_2H_2$ or saturated hydrocarbons such as $C_2H_6$, or even radicals originating from the dissociation of these gaseous products, as suspected by Gautier et al. (2014). Hydrocarbons do not act alone; gaseous nitrogen species such as nitrile $HC_3N$ or others must participate to explain the nitrogen signature already present in the first generation of aerosol analogues. The aliphatic character of the molecules making up aerosols at the start of their formation is also consistent with the fact that the atomic hydrogen abundantly released by the fragmentation of $CH_4$ or other hydrocarbons participates in a hydrogenation process at the surface of solid aerosols.

Subsequently, the solid growth observed up to 50s seems to be controlled more by chemical processes, where the spherical growth in volume of monomers is more restricted, compared to monomers produced with a longer growth time. However, we have observed that it is possible under these conditions to form spherical monomers with a micrometric size (without modifying the chemical composition of the monomers). As shown by Lavvas et al. (2011), the formation of these micrometric monomers may result from growth controlled by microphysical processes such as aggregation/coagulation, leading to the coagulation of a large number of solid monomers with each other (perhaps a hundred or so), which would have come from the first generation.

From 50 to 145 seconds, new growth mechanisms are observed, leading to new modifications in the properties of the monomers produced. The chemical composition of the solids continues to evolve. At around 50 seconds, hydrogen depletion has stabilized, with components presenting an H/C ratio fixed at around 1. Compared with the first generations of monomers, nitrogen enrichment has continued, with the N/C ratio doubling to 0.5, and majority compounds containing 6 nitrogen atoms. This increase in the number of unsaturations is also associated with a change in the structure of the majority compounds, which become aromatic. From then until 145 seconds, nitrogen incorporation continues slowly within the aromatic compounds. The morphology of the monomers shows a slowly increasing spherical volume, with an average retained diameter $d_{moy}$ stagnating around 400 - 500 nm.

From a duration of around 40 seconds, an acceleration in the kinetics of $CH_4$ gas consumption is observed in our experiment and induces more abundant ion production, also help by increasing the initial ion concentration by injecting an initial proportion of 20% $CH_4$, according to Wattieaux et al. (2015). As shown by Carrasco et al. (2012) and Gautier et al. (2014), the incorporation of nitrogen within organic solids is favoured by ionic chemistry processes and could explain the one we observe on aerosols that continued their growth after this acceleration of $CH_4$ dissociation. Starting with an initial 2% $CH_4$ diluted in $N_2$ gas, Berry et al. (2019b) simulated the organic gas chemistry of the titan atmosphere, photolyzing the $N_2$ - $CH_4$ mixture with EUV photons. By analyzing the gaseous chemical composition, Berry et al. (2019) identified the ions formed, a large majority of which are nitrogen-bearing, comprising mainly 1 to 4 nitrogen atoms. A participation of the ionic species detected in the experiments of Berry et al. (2019), may correlate with the incorporation of nitrogen observed in the evolution of the chemical composition of our aerosol analogues growing from 35 to 145 seconds. The evolution of solid components towards aromatic structures may also reflect the involvement of ionic chemistry on the surface of growing aerosols, which can break aliphatic C-H bonds, as shown by Imanaka et al. (2004). In addition, this evolution of the structure of solid constituents can affect the charge state of aerosols (through UV photoelectric ejection processes), and influence the size distribution of monomers, as well as solid coagulation processes as shown by Bakes et al. (2002) and Imanaka et al. (2004).

Indeed, microphysical coagulation of spherical monomers is observed, with participants having a similar diameter and a minimum of 300 - 400 nm. Between 50 and 145 seconds, this coagulation process preferentially takes place by combining a limited number of monomers (less than ten) and leads to the reformation of a new spherical monomer with a larger volume. This coagulation phenomenon may be favored by an average number of monomers present $N_{Part,moy}$, reaching and stagnating around the calculated maximum ($10^{+10}$), over this time period. As Lavvas et al. (2011) point out, monomer density influences the probability of collision between solids by increasing the rate of coagulation, but also the density of gaseous precursors present. During the coagulation process, growth chemistry on the surface of solid monomers through interaction with certain gaseous products can take place in parallel. This surface chemistry involves a loss of participating gases and can enable the coagulation of solid monomers to reform a spherical solid particle by smoothing the aggregate surface (Lavvas et al., 2011). Lavvas et al (2011) have considered the involvement of gaseous radicals such as $C_2H$, CN and HCCN in this surface chemistry, where their formation can be linked to the presence of neutral molecules such as $C_2H_2$ and HCN. In our experiments, before 50 seconds, neutral gaseous products show kinetics that characterize an increase in their quantities, while participating in the chemical processes of solid formation. When the abundance of gaseous products seems to reach a plateau (maximum value), the coagulation process of evolved monomers begins (from 50 seconds). Subsequently (between 45 and 70 seconds), the kinetics of gaseous products show a transition, where they begin a sub-phase characterizing a net consumption. This kinetic transition may have been influenced by the additional participation of gaseous products in surface chemical reactions with the monomers present, indicating the establishment of microphysical growth processes combined with surface chemical processes.

After 145 seconds and up to an unknown duration, the spherical growth in volume of the monomers increases sharply until reaching diameters on the order of ten micrometers. New aggregate structures containing around ten monomers with a diameter equivalent to 2 µm also appear. In addition, the monomers contained in these larger aggregates also exhibit coagulation processes leading to further volume growth of the monomers. In terms of the chemical composition of the solids, the constituents continue to become enriched in nitrogen with an N/C ratio migrating towards 1, and where in aromatic compounds containing N9 or N10 for example become among the majority classes. In parallel, we believe that the average number of monomers present within the plasma is certainly tending to decrease, perhaps testifying to the dominant involvement of microphysical processes of solid aerosol growth (coagulation) that favor the formation of large structures against monomer density, as shown by Lavvas et al. (2011).

During this final stage in the growth of Titan aerosol analogues, the concentrations of gaseous species become constant and stable from around 150 seconds. However during the transient regime, significant quantities have been consumed of gas such as HCN and $C_2H_2$ (around 75 - 90%). As described by Imanaka et al (2004), chemical processes characterizing nitrogen incorporation into aromatic rings can involve HCN via heterogeneous adsorption reactions on the surface of organic solids. Heterogeneous interactions between neutral gaseous species and monomers could correlate with the different gas kinetics and aerosol microphysics observed, giving solid aerosols the role of potential neutral gas wells within the plasma of the PAMPRE experiment.

**IV.b – Aerosol growth : comparison with modèle**

In this study, we observed a morphological and chemical time evolution during the growth of Titan analog aerosols, taking place under a fixed experimental condition in the PAMPRE dusty plasma reactor. According to our experimental observations, over time, different solid growth mechanisms may have succeeded one another, and been controlled by chemical or microphysical processes, each bringing different modifications to the physico-chemical properties of the solid monomers. In the following, we can now compare the growth mechanisms experimentally observed here on Titan analog aerosols, with the mechanisms described by the model of Lavvas et al. (2011) for Titan atmospheric aerosols.

To describe the stages in the formation and physical growth of Titan aerosols, Lavvas et al. (2011) sequenced this evolution according to three atmospheric regions: the upper atmosphere between altitudes of 1,000 and 650 km (ionosphere, thermosphere), the mesosphere between 650 and 500 km, and the lower atmosphere below 500 km (stratosphere, troposphere). In our experiments, we can also sequence the growth of Titan's analog aerosols according to three time periods: from 20 to 50s, from 50 to 145s, and after 145s, where the mechanisms occurring according to each time period have certain similarities with those considered to occur in each atmospheric region by Lavvas et al. (2011). Whether within an atmospheric region or within an experimental environment established over a period of time, the residence time of solids also characterizes that of their growth, indirectly controlling its development by the reactions imposed and limited as a function of the environment and the length of time the aerosols remain there.

The scenario of Lavvas et al. (2011) predicts that at around 1000 km altitude, polymerization of gaseous benzene with radicals ($C_2H$, CN, HCCN) produces polycyclic aromatic polymers (PAH, nitrogen-containing PAH) until the first solid monomers with a spherical shape and nanometer diameter are generated. In this model, following the appearance of the first monomers, their sedimentation rate is rapid and will impose a short residence time for them at high altitudes, still allowing growth to begin by chemical processes up to an altitude of 650 km. This chemical growth mechanism will induce a carbonaceous complexification of the chemical composition of the solids through an increase in the average number of C carbon atoms, and the model considers that the monomers retain a spherical shape with a slow increase in the average diameter remaining below ten nanometers at 650 km.

In our experimental simulation between 20 and 50 seconds, the formation of the first solid monomers is achieved by nucleation involving several copolymers such as neutrals like $C_2H_2$, HCN, $HC_3N$, $C_2H_6$, or radicals from the photolysis of these gaseous products. These first analogous monomers are chemically made up of compounds possessing mostly 3 or 4 nitrogen atoms with an aliphatic structure, and whose morphology has a quasi-spherical shape with a diameter on the order of a hundred nanometers. Up to 50 seconds, we observe a nitrogenous complexification of the monomers chemical composition, with compounds containing 1 additional nitrogen atom (5 or 6) becoming dominant. During this first observed chemical growth, analogous monomers retain a spherical morphology with an average diameter increasing by a factor of 2 but remaining below 400 nm. We note that the formation of multi-centi-nanometric monomers may be due to the aggregation of a large number of solid monomers possessing a smaller size and capable of orienting into a radial structure, as observed in previous experimental studies (Hadamcik et al., 2009; Sciamma'O-Brien et al., 2017).

During this first phase of aerosol analogues growth, we observe in agreement with Lavvas et al. (2011), that the dominant processes are chemical. However, while Lavvas et al. (2011) suggested that these processes were regulated by the chemistry of carbon with aromatic solid compounds we observe experimentally that these early chemical processes of growth appear to be regulated by the reactivity of nitrogen with branched (aliphatic) solid constituents.

In Titan's atmosphere between 650 and 500 km, Lavvas et al. (2011) consider that a transition is taking place in the mechanisms that predominate aerosol growth, similar to what we observe experimentally during the second growth period of our analogues. Whether through an increase in the number of monomers produced within experiments, or through an accumulation of atmospheric aerosols by decreasing their sedimentation rates in Titan's mesosphere, environments start to become dominated by the presence of monomers, which favors collision between solids, leading to monomer coagulation and the formation of aggregate structures. Lavvas et al (2011) assume that only the coagulation of solid monomers affects their shape and size, i.e. when growth becomes controlled by microphysical laws/processes. The formation of aggregates will induce: morphological structures with a shape that differs from a sphere; an increase in the average surface area represented by the solids, leading to an increase in the probability (rate) of collision between aerosols as well as their probability of coagulation. According to this model, up to 500 km the number of monomers within an aggregate can reach a hundred, inducing a decrease in aerosol density but also a decrease in the density of gaseous precursors considered for solids. They also predict that microphysical aggregation of monomers can be accompanied by surface chemical growth of the solids by direct reaction with gaseous radicals, leading to an increase in the proper diameter of the monomers in the aggregates.

Similarly, between 50 and 145 s in our experimental simulation, the appearance of aggregates of evolved monomers is observed, whose structure no longer represents a sphere. We note that these microphysical structures contain a limited number of monomers (less than ten), compared with aggregates observed later in the experiments (after 145s). With regard to the range of diameters measured on monomers in aggregates or not, we observe that this increases slowly, with an average diameter stagnating at around 400 - 500 nm. In parallel, analogous monomers show a chemical evolution, where heavier constituents with an aromatic structure form and replace the initial aliphatic constituents. Up to 145s, nitrogen incorporation continues within solid aerosols, forming compounds with a majority of 6 or more nitrogen atoms. As shown by Bakes et al (2002) and Imanaka et al (2004), the formation of solid aromatic constituents can affect the charge state of analog aerosols, and thus strongly influence the coagulation process of monomers by controlling their size distribution. According to our experimental observations carried out between 50s and 145s, solid monomers that have undergone chemical evolution (presence of aromatic nitrogen compounds) are those coagulating to form non-spherical aggregates. We observe that the coagulation process takes place until the monomers coagulate together to reform a single smooth monomer that is once again spherical with a larger diameter, but whose increase in diameter remains low. We also suspect that surface chemistry begins in parallel, perhaps influencing coagulation, and also favoring the surface growth observed on our analogous monomers produced after 145s in our experiments.

In the stratospheric region that is most enriched in aerosols on Titan (below 500 km), Lavvas et al. (2011) predict that a local increase in aerosol density promotes the coagulation process contributing to rapid growth of the surface represented by solids. Simultaneously, these conditions also promote surface reaction rates, until surface chemistry becomes dominant. Surface chemistry is also expected to influence the shape of the aggregates, where they suspect that the numerous direct reactions taking place at the surface of the solids with gaseous radicals in their model, can lead to coagulation of the monomers to re-form a sphere that can reach volumes of the micrometer order (surface smoothing effect). These surface chemical reactions can be defined simply as the bonding efficiency of a gaseous component to a solid surface and characterized by a coefficient describing the probability of this interaction for a specific gas and solid. For this chemical growth process, Lavvas et al (2011) take particular account of the participation of the HCCN radical, assigning the maximum possible value to the uptake coefficient (equal to 1).

In our experiments, for longer durations (greater than 145s), we observed the formation of spherical monomers with diameters ranging from a few micrometers to around 15 µm. We also observed the formation of new aggregate structures comprising more than a dozen spherical monomers, but with an average diameter of several micrometers. Monomers included in an aggregate also exhibit growth mechanisms by coagulation linked to surface chemistry. The most advanced monomer analogues in our experiments feature a chemical composition in which the aromatic compounds include additional nitrogen incorporation, compared with analogues produced up to a duration of 145s.

These experimental morphological observations correlate with the involvement of surface chemistry in Titan analogues, as described by Lavvas et al. (2011). We have estimated uptake coefficient values for some neutral gaseous products. Indeed, during our experiments, the kinetics observed for some neutral gases show a slope that can be qualified as net consumption (loss). At different times during the experiment, the calculated number of monomers present and the measured dimensions of the monomers made it possible to determine the total surface area represented by the solids present, and thus to relate this to the rate of gas loss to determine a uptake coefficient. The calculated values of the uptake coefficient, considered as an upper limit, oscillate between $10^{-4}$ - $10^{-5}$ for the neutral gases considered, such as $C_2H_2$ and HCN, and the solid surface of our Titan analog monomers.

**V - Conclusion :**

**Acknowledgements:**
Nathalie Carrasco thanks the European Research Council for funding via the ERC OxyPlanets projects (grant agreement No. 101053033). Thomas Gautier acknowledges funding by the Agence Nationale de la Recherche under the grant ANR-20-CE49-0004-01.